\documentclass[11pt,a4paper]{article}
\pdfoutput=1
\usepackage{jheppub}
\usepackage{amsmath,amssymb,enumerate,graphicx,caption,subcaption,fancyhdr,tikz}
\usepackage[T1]{fontenc}
\usetikzlibrary{positioning}

\newcommand{\ud }{\mathrm{d}}

\newcommand{\figref}[1]{Figure \ref{#1}}
\newcommand{\secref}[1]{Section \ref{#1}}
\newcommand{\cincludegraphics}{\centering\includegraphics}

\title{Moduli Space Dynamics of Noncommutative U(2) Instantons}

\author{Andrew Iskauskas,}
\author{Douglas J. Smith}

\affiliation{Department of Mathematical Sciences, Durham University \\
Lower Mountjoy, Stockton Road, Durham DH1 3LE, UK}

\emailAdd{andrew.iskauskas@durham.ac.uk}
\emailAdd{douglas.smith@durham.ac.uk}

\abstract{We consider the low energy dynamics of charge two instantons on noncommutative $\mathbb{R}^{2}_{NC}\times\mathbb{R}^{2}_{NC}$
in U(2) 5-dimensional super-Yang-Mills, using the Manton approximation for slow-moving instantons to calculate the moduli space metric. By employing
the ADHM construction, we are able to understand some aspects of the geometry and topology of the system. We also consider the effect of adding a potential to the moduli space, giving scattering results for noncommutative dyonic instantons.}

\keywords{instantons, noncommutative, ADHM, Yang-Mills}

\begin{document}

\maketitle
\flushbottom

\section{Introduction}
Instantons have long provided a fertile testing ground for exploring aspects of Yang-Mills theory \cite{Belavin:1975}, and can play an important role in determining the behaviour of non-perturbative effects in supersymmetric Yang-Mills theory \cite{Belitsky:2000, Dorey:2002, TongTASI:2005}. In comparison with other solitons, however, little is known about their dynamics. In particular, when compared to monopoles, this paucity of information is most apparent. \\
Instantons naturally arise in the study of 4-dimensional Euclidean Yang-Mills, in which there exists no dynamical structure. However, it is also possible to embed such solutions in higher-dimensional Yang-Mills theories, in which a time component can be explicitly considered. The study of $5$d SYM is of great interest, as in this context the instantons appear as particles of the theory. Indeed, the instantons appear as $1/2$-BPS states which, in the low-energy limit of the theory, correspond to D0-branes dissolved in a system of D4s \cite{Witten:1996}. The previously considered 4-dimensional Euclidean instantons may be lifted to this $4+1$ dimensional theory by identifying them with the static solutions of theory.

There are further motivations for the study of instanton systems. It has been conjectured that the theory of coincident D4-branes is dual to the compactified theory of coincident M5-branes on $\mathbb{R}^{4,1}\times S^{1}$, and is UV complete with the addition of instantons \cite{Lambert:2011, Douglas:2011}. Instantons play a key role in matching the BPS spectrum of these theories, where the instanton charge corresponds to the Kaluza-Klein momentum associated to the compactified M-theory. The index of the degenerate BPS states can be calculated from localisation techniques as in \cite{Stern:2000}, and the same results were obtained for the $\mathcal{N}=8$ QM theory from the moduli space of a single U$(N)$ instanton in \cite{Bak:2013}.

There are also a large number of identifications that can be made between instantons and other solitonic solutions in reduced dimensions. It is known that noncommutative instantons in SU$(2N)$ displaying SO$(3)$ invariance can produce a class of non-Abelian vortices \cite{Eto:2012, Hanany:2003}; by considering instantons whose ADHM data has circle invariance, one can obtain monopoles in hyperbolic space with platonic symmetries \cite{Manton:2014, Cockburn:2014}; and it is believed that a more general class of vortices coupled non-trivially to a gauge field can be obtained by considering a dimensional reduction of noncommutative dyonic instantons \cite{Samols:1992}. As an extension, a large body of material is dedicated to the study of vortex systems with impurities, thus providing an entry point into problems considered in condensed matter physics: see, for example, \cite{Cho:2003}.

It is not straightforward to gain a deep understanding of the dynamics of instantons on the full field theory. Instead, it proves fruitful to employ an observation due to Manton \cite{Manton:1982} and study the motion of instantons as geodesics on the moduli space of solutions. The moduli space is a $4kN$-dimensional space made up of all instanton solutions for a given gauge field SU$(N)$ and topological charge $k$. Configurations within this moduli space can be seen as minimum energy solutions of the field theory and, should we perturb such a solution by a small velocity, we expect that it will remain in (or energetically close to) the moduli space. it transpires that it is possible to view the dynamics of slow-moving instantons as geodesic motion on this moduli space endowed with a suitable hyperK\"ahler metric, and it then becomes feasible to consider low-energy scattering and evolution of the field theory.

Recall that in the D4-brane theory we have $5$ transverse scalar fields, valued in the adjoint of the gauge group SU$(N)$, which describe the positions of the branes in the space transverse to the worldvolume. If we choose to separate the branes, and give any of these scalars a non-zero expectation value, then we introduce a non-zero scalar field into the instanton equations. The minimum-energy solutions become $1/4$-BPS and the $A_{0}$ component of the gauge field becomes proportional to this non-zero Higgs field  \cite{Lambert:1999,Dorey:2002}. This imbues the instantons with an electric charge which balances the effect of the Higgs field. In terms of the moduli space, this addition produces a non-zero potential \cite{Lambert:1999, Peeters:2001}. From the point of view of the underlying string theory, these dyonic instantons form a bound state of fundamental strings and D0-branes.

The moduli space of instantons constructed contains singularities arising from instantons of zero size. Such ``small'' instantons have a dual picture in the string theory of a transition between the Higgs and Coulomb branches of the D$0$ theory \cite{TongTASI:2005}. The Coulomb branch of the theory corresponds to D0 branes separated from the D4s: the moduli describe the positions of the D0s transverse to the D4s. The Higgs branch corresponds to the D0s `dissolved' in the D4s, and their moduli are precisely the moduli of instantons in the Yang-Mills theory. The singularity in the metric of this moduli space, attained when the instantons hit zero-size, then corresponds to the transition point between the two branches. To circumvent this problem, it is possible to use a noncommutative framework in which a minimum bound is placed on the instantons' size via the introduction of a Fayet-Iliopoulos term \cite{Nekrasov:1998}. This modification to the theory smooths out the moduli space singularities, and it has been seen explicitly that the metric takes Eguchi-Hanson form in the case of a single U$(1)$ instanton \cite{TongLee:2001}. The ADHM procedure applied to a noncommutative system returns the expected results: namely, solutions are self-dual and maintain integer charge \cite{Chu:2002}.

The dynamics of the commutative dyonic instanton with gauge group SU$(2)$ have been studied for a single instanton and two well-separated instantons. The (super-)metric of the moduli space of $2$ commutative instantons in the absence of a potential was studied via the hyperK\"ahler construction in \cite{Bruzzo:2001} and with the addition of a potential in \cite{Peeters:2001}. More recently an extensive analysis of the dynamics have been studied for two instantons with arbitrary separation \cite{Allen:2013}. A free single instanton may evolve into a configuration where its size $\rho$ and hence its angular momentum $\rho^{2}\dot{\theta}$ can vanish, resulting in the small instanton. The introduction of a potential term guarantees that this singular point can not be reached for a single instanton that starts with a non-zero angular momentum and a bounded, non-zero, size. Specifically, it will remain in a stable orbit with conserved angular momentum. In the case of multiple instantons, however, this may not hold: the instantons may trade angular momentum with each other, allowing one instanton to grow in size at the expense of its counterpart, approaching the zero-size singularity in finite time. This was shown in \cite{Allen:2013}. The zero-size singularity still exists, therefore, for more than a single dyonic instanton; we must consider a noncommutative deformation to the space in order to remove the singularity.

The outline of this paper is as follows. In \secref{sec:Preliminaries}, we review the construction of instanton solutions as solutions to the self-dual Yang-Mills field equations. A consideration of solitonic solutions, via the Bogomolny argument, leads one naturally to an algebraic formulation of instantons for a given topological charge, $k$. The results extend to noncommutative spaces; we summarise the connection between noncommutative function space and the quantum mechanical analogue. Having constructed solutions, we consider the parameter space of the charge $k$ instantons as furnishing a moduli space of allowed configurations, and may derive an algebraic formalism for determining the metric on this moduli space. This allows us to analyse the dynamics of two instantons via the Manton approximation \cite{Manton:1982}. Finally, we consider the effect of introducing a non-zero electric charge, or potential, on the moduli space.\\
In \secref{sec:Construction}, we proceed to explicitly derive the solutions for $2$ U($2$) instantons in both the commutative and noncommutative frameworks. The presence of noncommutativity perturbs the known solutions in a non-trivial manner, and by finding an expedient parametrisation for this perturbation we may calculate the metric of the noncommutative $2$-instanton system. Due to the induced complexity of solutions, it is not easy to find a description of the full, $16$-dimensional moduli space. However, we may make use of some global symmetries of the system to consider a geodesic submanifold of this space. With this reduction, explicit results may be obtained. We consider the results and, as expected, we find that the manifold generated is in fact smooth and singularity-free, unlike in the commutative case. This is indicative of the results gained in \cite{TongLee:2001} where the single instanton moduli space was seen to correspond to the Eguchi-Hanson metric, which contains no orbifold singularities.\\
In \secref{sec:Scattering}, we use the results gained to consider dynamics, and in particular scattering, of the two noncommutative instantons. The presence of a non-zero Fayet-Iliopoulos term in the overarching field theory has profound consequences for the results gained: most strikingly, right-angled scattering (a distinguishing feature of most soliton dynamics) is no longer the natural behaviour, even for a vanishing Higgs field. In fact, a wide range of behaviours are present, of which scattering at $\pi/2$ is only one possible outcome.\\
In \secref{sec:Dyonic}, we extend the analysis of the previous section to the dyonic instantons. The results obtained herein suggest that one may consider the noncommutativity to function as an ersatz effective potential on the moduli space of commutative instantons. The dynamics of two commutative instantons admits orbiting solutions, where the attractive force of the potential is balanced by the natural repulsive force of the instantons. In the noncommutative picture, we find an analogous result, with some interesting modifications: previously stable orbiting configurations can become unstable in finite time, demonstrating scattering or attraction, with varying noncommutative strength. \\
Finally, in \secref{sec:Summary} we summarise our results and consider extensions to the present work.

\section{The construction of instantons}\label{sec:Preliminaries}
In this section, we review instantons in $(4+1)$-dimensional Yang-Mills theory. This will encompass both `free', $1/2$-BPS, instantons and their dyonic $1/4$-BPS counterparts. We briefly outline the ADHM construction for such a field theory, and the connection between the free parameters therein with collective coordinates on a moduli space. We proceed to consider the key differences between the commutative and noncommutative formulations.

\subsection{Instantons in $5$d Yang-Mills}
We first consider the underlying string theoretical interpretation of Yang-Mills theory. The low-energy dynamics of a stack of $N$ coincident D$4$-branes may be identified with a U($N$) super-Yang-Mills field theory \cite{Zwiebach:2004}. Such a system preserves one half of the supercharges, and is thus described by an $\mathcal{N}=2$ SUSY theory in five dimensions. Open strings stretched between the D$4$-branes give rise to a U($N$) world-volume gauge symmetry, with associated gauge field $A_{\mu}$, $\mu=0,1,\dots,4$. The theory also contains $5$ adjoint scalars $X^{I}$, $I=5,6,\dots,9$, describing the branes' relative positions in the transverse directions. By factoring out the centre of mass from the theory we obtain $5$-dimensional super-Yang-Mills.

For the purposes of considering instantons, we henceforth consider only the bosonic sector of the theory, with (for convenience) gauge coupling set to one. The associated action is
\begin{equation}
S=-\int \ud^{5}x\text{Tr}\left(\frac{1}{4}F_{\mu\nu}F^{\mu\nu}+\frac{1}{2}D_{\mu}X^{I}D^{\mu}X^{I}+\frac{1}{4}[X^{I},\,X^{J}]^{2}\right),
\end{equation}
where the covariant derivative is given in standard form
\begin{equation*}
D_{\mu}X^{I}=\partial_{\mu}X^{I}-i[A_{\mu},\,X^{I}]
\end{equation*}
and the field strength is
\begin{equation*}
F_{\mu\nu}=\partial_{[\mu}A_{\nu]}-i[A_{\mu},\,A_{\nu}].
\end{equation*}
While the construction of instantons is valid for all choices of gauge group U($N$), the calculational complexity vastly increases with larger gauge groups. We consider only a stack of two D$4$-branes, so that the gauge group is U$(2)$. In the commutative case, the U$(1)$ factor decouples from the theory; as we will discuss, this is not true in the noncommutative picture. As well as the world-volume and transverse indices outlined above, we will also use the indices $i,\,j$ to denote the purely spatial directions of the $5$d theory.

We may assume that just one of the transverse scalar fields $X^{5}\equiv\phi$ is non-zero. The induced Higgs VEV, $\langle\phi\rangle$, will correspond to the separation of the branes in the $X^{5}$ direction. This is equivalent to any other choice of transverse brane separation up to some SO($5$) rotation of the $X^{I}$, and in choosing a particular direction we break the $R$-symmetry of the full Yang-Mills theory. However, this does not affect the validity of the analysis (and, in fact, is crucial in certain identifications with lower-dimensional solitonic theories). The energy of the system is
\begin{equation}
E=\int\ud^{4}x\text{Tr}\left(\frac{1}{2}F_{i0}F_{i0}+\frac{1}{4}F_{ij}F_{ij}+\frac{1}{2}D_{0}\phi D_{0}\phi+\frac{1}{2}D_{i}\phi D_{i} \phi \right).
\end{equation}
In order to obtain solitonic solutions, we seek to find minimum energy solutions. The requirement for finite energy solutions is satisfied by demanding that the gauge field becomes pure gauge at spatial infinity: that is
\begin{equation*}
A_{i}=-\partial_{i}g^{\infty}(g^{\infty})^{-1}
\end{equation*}
as $|x|\rightarrow\infty$. The map $g^{\infty}:\, S^{3}_{\infty}\rightarrow \text{U}(2)$ defines a winding number from the sphere at infinity to the gauge group, the degree of which is given by the second Chern number $c_{2}\in\mathbb{Z}$. We define, for identification, the following quantities:
\begin{align*}
k&\equiv-\frac{1}{8\pi^{2}}\int\ud^{4}x\epsilon_{ijkl}\text{Tr}(F_{ij}F_{kl}), \\
\mathcal{Q}_{E}&\equiv\int\ud^{4}x\text{Tr}(D_{i}\phi F_{i0}).
\end{align*}
These are to be interpreted as the topological charge and electric charge, respectively, of the theory. The topological charge is equivalent to the winding number, and so for a given $k\in\mathbb{Z}$ we may consider the family of all instantons with winding $k$. Such solutions may smoothly deform into one another, but must remain in this $k$-sector of the theory. Hence, in the instanton description, each successive value of $k$ decouples from all others and this will allow us to consider evolution and scattering of $k$-instanton solutions for a particular $k$ \cite{Allen:2013}.

Employing the standard Bogomolny argument \cite{Bogomolny:1975} to bound the energy, we find
\begin{align*}
E=\int\ud^{4}x\text{Tr}\Bigg(&\frac{1}{4}(F_{ij}\pm\star F_{ij})^{2}\mp\frac{1}{2}\, F_{ij}\star F_{ij} \\
&+\frac{1}{2}(F_{i0}\pm D_{i}\phi)^{2}\mp F_{i0}D_{i}\phi+\frac{1}{2}D_{0}\phi^{2}\Bigg),
\end{align*}
where $\star F_{ij}\equiv\frac{1}{2}\epsilon_{ijkl}F_{ijkl}$ is the Hodge dual of the field strength. The choices of sign in this expression are correlated within each line, but independent between the two lines. Then the energy is bounded by
\begin{equation*}
E\ge2\pi^{2}|k|+|\mathcal{Q}_{E}|,
\end{equation*}
and this Bogomolny bound is saturated when
\begin{align}\label{eq:Bogomolny}
\begin{split}
F_{ij}&=\pm\star F_{ij}, \\
F_{i0}&=\pm D_{i}\phi, \\
D_{0}\phi&=0.
\end{split}
\end{align}
These are the BPS equations for dyonic U($2$) instantons. The first equation requires that the field strength be (anti-)self-dual, and the second and third are satisfied when the fields are static and $A_{0}=\pm\phi$. Since each $k$-sector decouples from all others we need only consider either the self-dual or anti-self-dual case, which we denote as instantons or anti-instantons respectively. We henceforth consider only the self-dual case, yielding $k$-instantons. It will still be necessary to satisfy the background field equations for the scalar field, namely
\begin{equation}\label{eq:BGField}
D^{2}\phi=0.
\end{equation}
This requirement will be important in the consideration of dyonic instantons.

The Bogomolny equations, while simpler than those of the full Yang-Mills theory, do not trivially admit analytic solutions. Fortunately, the ADHM construction \cite{ADHM:1978} relates these differential constraints on the gauge field to purely algebraic ones. This will allow us to explicitly construct classes of self-dual instantons whose induced gauge field automatically satisfies the Bogomolny equations \eqref{eq:Bogomolny}. Before we apply the ADHM construction, however, we consider the noncommutative analogue.

\subsection{Noncommutative $\mathbb{R}^{4}$}\label{subsec:Noncomm}
As described above, the study of instantons allows us to find non-trivial solutions to the Yang-Mills field equations in (static) Euclidean $\mathbb{R}^{4}$ that would otherwise be occluded. In the previous section, the spatial $\mathbb{R}^{4}$ (consisting of $x_{i}$, $i=1,\dots,4$) admits trivial commutation relations between each direction. For reasons that shall become apparent, we may introduce an underlying noncommutative geometry to the theory by making some, or all, of these commutation relations non-zero. This is equivalent to choosing a preferred complex structure on the space. We stipulate the following commutation relations:
\begin{equation}\label{eq:Commutations}
[x_{i},\,x_{j}]=i\theta_{ij}
\end{equation}
where $\theta_{ij}$ is a real, antisymmetric matrix. Without loss of generality, we may break the underlying SO$(4)$ symmetry of the space and express $\theta$ in a simpler form \cite{TongLee:2001}
\begin{equation}\label{eq:CanonicalTheta}
(\theta_{ij})=\left(\begin{matrix} 0 & -\theta_{1} & 0 & 0 \\ \theta_{1} & 0 & 0 & 0 \\ 0 & 0 & 0 & -\theta_{2} \\ 0 & 0 & \theta_{2} & 0 \end{matrix} \right)
\end{equation}
for $\theta_{1}$ and $\theta_{2}$ real. Classically, if both of the $\theta_{i}$ are non-zero then we may scale the two coordinate directions corresponding to, say, $\theta_{1}$ such that the noncommutative parameters have equal magnitude. The condition that $\theta_{ij}$ is self- or anti-self-dual is equivalent to requiring that $\theta_{1}-\theta_{2}=0$ or $\theta_{1}+\theta_{2}=0$, respectively. We shall examine the difference between the two cases shortly.

From the perspective of the Yang-Mills field theory, the introduction of a noncommutative background induces a deformation in the notion of multiplication: one now must consider functions multiplied using the Moyal-$\star$ product. For functions $f(x)$ and $g(x)$ valued in $\mathbb{R}^{4}_{NC}$, we have
\begin{equation}\label{eq:Moyal}
f\star g(x)=\left.\text{exp}\left(\frac{i}{2}\theta_{ij}\partial_{i}\partial_{j}^{\prime}\right) f(x)g(x^{\prime})\right|_{x^{\prime}=x}.
\end{equation}
This gives an expansion in powers of $\theta$:
\begin{equation*}
f\star g(x)=f(x)g(x)+\frac{i}{2}\theta^{ij}\partial_{i}f(x)\partial_{j}g(x)+\mathcal{O}(\theta^{2}).
\end{equation*}
In this noncommutative framework, the gauge field $A_{i}$ transforms as
\begin{equation*}
A_{i}\mapsto g^{-1}\star A_{i}\star g+g^{-1}\star\partial_{i}g,
\end{equation*}
where $g$ takes values in U($N$). The field strength is correspondingly adjusted as
\begin{equation*}
F_{ij}=\partial_{[i}A_{j]}-i[A_{i},\,A_{j}]_{\star},
\end{equation*}
where we denote the commutator with $\star$ to emphasise the non-standard multiplication therein.

From the point of view of finding solutions to the Bogomolny equations \eqref{eq:Bogomolny}, working in the noncommutative framework allows for a greater range of instanton configurations, circumventing Derrick's theorem due to the additional length scale $[\theta_{i}]=\text{length}^{2}$. However, with the above formalism, one would have to calculate such solutions to all orders in $\theta$ which (with the exception of the simplest cases) is severely non-trivial and prevents any meaningful analysis. We may proceed due to an isomorphism between the algebra of functions with the $\star$-product and the algebra of operators on some Hilbert space, as demonstrated in \cite{Gopakumar:2000}. This identification will allow us to utilise the ADHM procedure in the noncommutative framework.

Consider, for simplicity, a noncommutative theory in $\mathbb{R}^{2}$, giving a single non-trivial spatial commutation relation $[x_{1},\,x_{2}]=i\theta_{12}$. Then for a generic function on this space, we have the associated operator on the space of Hilbert functions of the analogous quantum system
\begin{equation}
\hat{\mathcal{O}}_{f}(\hat{x}_{1},\,\hat{x}_{2})=\frac{1}{(2\pi)^{2}}\int\ud^{2}\alpha \,U(\alpha_{1},\,\alpha_{2})\tilde{f}(\alpha_{1},\,\alpha_{2})
\end{equation}
where $\tilde{f}(\alpha_{1},\,\alpha_{2})$ is the Fourier transform of $f(\hat{x}_{1},\,\hat{x}_{2})$, and $U(\alpha_{1},\,\alpha_{2})=\exp\left(-i(\alpha_{1}\hat{x}_{1}+\alpha_{2}\hat{x}_{2})\right)$. We may now seek an expression for $\hat{\mathcal{O}}_{f}\hat{\mathcal{O}}_{g}$: using Baker-Campbell-Hausdorff and a suitable change of variables, we find
\begin{align*}
\hat{\mathcal{O}}_{f}\hat{\mathcal{O}}_{g}=&\frac{1}{(2\pi)^{2}}\int\ud^{2}\gamma\,U(\gamma_{1},\,\gamma_{2})\widetilde{f\star g}(\gamma_{1},\,\gamma_{2}) \\
 =&\hat{\mathcal{O}}_{f\star g}.
\end{align*}
This shows that the Moyal $\star$-composition of two functions on a commutative space has a direct analogue in the composition of operators in a Hilbert space. Given this correspondence, we may derive spatial commutation relations of functions on $\mathbb{R}^{2}_{\text{NC}}$ as operator relations on a Hilbert space of operators, and vice versa. This correspondence is key to a consistent definition of the ADHM operators in noncommutative scenarios.

With these considerations, the ADHM procedure can be seen to follow in precisely the same manner as the commutative analogue but for the fact that the underlying gauge group of the gauge field is U($2$), rather than SU($2$) (the term $A^{4}_{i}1_{2}$ can be considered in the commutative case, but decouples from the theory and therefore has no impact on the analysis). This stems from the fact that the `simpler' gauge group SU($2$) is not closed under the Moyal $\star$-product multiplication \cite{Hashimoto:1999}. Explicitly, we have
\begin{equation*}
A_{i}=A_{i}^{a}\frac{\sigma^{a}}{2}+A_{i}^{4}\frac{1_{2}}{2},
\end{equation*}
where $\sigma^{a}$ provide the normal Pauli matrix representation of SU($2$). This isomorphism validates the use of the ADHM toolbox, to which we now turn, in a noncommutative framework.

\subsection{The ADHM Construction}\label{subsec:ADHM}
The ADHM construction allows a class of algebraic constraints to be explcitly formulated for a given instanton number $k$ and gauge group U$(N)$. The subject is well-documented (See, for example \cite{ADHM:1978,Allen:2013,MantonSutcliffe:2004}), and we will not reiterate the details. Formally, the data $\Delta$ is a $(2k+N)\times2k$ complex-valued matrix, up to some constraints, whose free parameters form a $4kN$-dimensional moduli space of allowed instanton configurations. Given the ADHM requirement
\begin{equation}
\Delta^{\dagger}\Delta=1_{2}\otimes f^{-1}(x)
\end{equation}
for $f^{-1}(x)$ invertible and $\Delta$ in `canonical form':
\begin{equation*}
\Delta(x)=a-bx\equiv\begin{pmatrix} L \\ M \end{pmatrix}-x\begin{pmatrix} 0 \\ 1_{2k} \end{pmatrix}
\end{equation*}
then the ADHM constraints on the moduli space parameters are
\begin{equation}\label{eq:ADHMConstraint}
L^{\dagger}L+M^{\dagger}M+\bar{x}x=1_{2}\otimes\tilde{f}^{-1}(x).
\end{equation}
Having solved these constraints, one may define a normalisable vector $U$ in the null space of $\Delta$, and derive the gauge field $A_{i}=iU^{\dagger}\partial_{i}U$. This procedure will guarantee a (anti-)self-dual field strength tensor $F_{ij}$. Up to local gauge transformations of the ADHM data, one may use this result to find a metric on the moduli space of charge $k$ U$(N)$ instantons. Denoting the free parameters in the ADHM data as $z_{r}$, for $r=1,2,\dots,4kN$, we obtain
\begin{equation*}
S=\frac{1}{2}\int\ud t\,g_{rs}\dot{z}^{r}\dot{z}^{s}
\end{equation*}
where
\begin{equation*}
g_{rs}=\int\ud^{4}x\,\text{Tr}(\delta_{r}A_{i}\delta_{s}A_{i})
\end{equation*}
and $\delta_{r}A_{i}$ are the zero modes of the space, namely gauge-invariant variations of the field $A_{i}$ in a direction that does not change the energy of the field configuration. This procedure corresponds, from the perspective of the D-branes, to gauge-fixing away the local U$(N)$ gauge transfomations, which do not act on physical states. Explicitly, the zero-modes may be written as
\begin{equation*}
\delta_{r}A_{i}=\partial_{r}A_{i}-D_{i}\epsilon_{r},
\end{equation*}
where $\epsilon_{r}$ is chosen such that Gauss' law is satisfied and $D_{i}\delta_{r}A_{i}=0$. Geodesic motion on this space, provided the velocity of the instantons is suitably small \cite{Manton:1982}, can be seen as equivalent to evolution of instanton configurations of a given charge $k$. This metric can be calculated via an explicit derivation and classification of the allowed zero-modes \cite{TongTASI:2005}, or via the ADHM data itself \cite{Osborn:1981}. In the same manner, a potential can be introduced to the space (representing a Higgs VEV separating the D-branes in the string theory picture) whose derivation in terms of the moduli space proceeds in a similar spirit to the `free' metric \cite{Allen:2013,Bak:2013} with the additional requirement of satisfying the Yang-Mills background field equation, $D^{2}\Phi=0$.

There are a number of difficulties to overcome in moving to the noncommutative picture. As mentioned earlier, the choice of noncommutative background affects the solutions we may obtain: as shown in \cite{Chu:2002}, should one choose a background with the `same' duality as the Yang-Mills solutions, the completeness relation $\Delta f\Delta^{\dagger}+UU^{\dagger}=1_{4}$ will not hold. Concretely, a search for solutions to $F_{ij}=\star F_{ij}$ in $\mathbb{R}_{NC}^{4}$ where $\theta_{1}=\theta_{2}$ will either result in solutions for $\Delta$ and a non-normalisable $U$, or a normalised $U$ with inconsistent $\Delta$. Consequently, we limit our search to self-dual instantons, that is those with topological charge $k>0$, and work in anti-self-dual $\mathbb{R}_{NC}$ such that $\theta_{1}=-\theta_{2}\equiv\zeta$ for $\zeta>0$. Note that we need not worry about this restriction as the choice is equivalent, from the point of view of the Yang-Mills theory, to considering anti-instantons on a self-dual $\mathbb{R}^{4}_{NC}$.

Secondly, the ADHM constraints themselves are no longer as simple. Representing $\Delta$ as matrix of quaternions, then the term $\bar{x}x$ in \eqref{eq:ADHMConstraint} is automatically proportional to the identity in the commutative case and so does not have an effect on the solution for $L$ and $M$. In the noncommutative case, using the commutation relations \eqref{eq:Commutations} with $z_{1}=x_{2}+ix_{1}$ and $z_{2}=x_{4}+ix_{3}$, the relevant expression is instead
\begin{equation}
\begin{aligned}
\bar{x}x&=\begin{pmatrix} \bar{z}_{2} & -z_{1} \\ \bar{z}_{1} & z_{2} \end{pmatrix}\begin{pmatrix} z_{2} & z_{1} \\ -\bar{z}_{1} & \bar{z}_{2} \end{pmatrix} \\
&=(\bar{z}_{1}z_{1}+\bar{z}_{2}z_{2})1_{2}+\begin{pmatrix} -2\zeta & 0 \\ 0 & 2\zeta \end{pmatrix},
\end{aligned}
\end{equation}
where the first term on the final line is the commutative result. The additional piece must be absorbed into the solution for $L$ and $M$ in a suitable manner.

Finally, we remarked in \secref{subsec:Noncomm} that the introduction of noncommutativity forces us to consider the full U$(2)$ gauge group, as the U$(1)$ piece is no longer frozen out. This modifies the construction of the metric on the moduli space in a non-trivial manner, via the global symmetries of $\Delta$. This, too, is surmountable, as shall become clear.

\section{Noncommutative U($2$) instantons}\label{sec:Construction}
In this section, we turn our attention to finding explicit solutions to the noncommutative ADHM constraints for two U($2$) instantons. This will allow us to generate the moduli space metric, consider scattering, and analyse the symmetries of the data. While to consider geodesics on the moduli space of the full data (comprising $16$ free parameters) is computationally expensive, we may use the symmetries inherent in the metric to consider geodesic submanifolds of the moduli space.

\subsection{The commutative $k=2$ data}
We first record, for comparison, the commutative $k=2$ data presented in \cite{Allen:2013}. The blocks of $\Delta$ are written explicitly in terms of quaternions as
\begin{align*}
L&=\left(\begin{matrix} v_{1} & v_{2} \end{matrix}\right), \\
M&=\left(\begin{matrix} \tau & \sigma \\ \sigma & -\tau \end{matrix}\right),
\end{align*}
which satisfy the symmetry requirements of the ADHM constraints
\begin{equation*}
\Delta^{\dagger}\Delta=1_{2}\otimes f^{-1}(x).
\end{equation*}
The remainder of the ADHM constraints, namely $a^{\dagger}a=\mu^{-1}1_{2}$, split into two parts. The diagonal elements yield $|v_{1}|^{2}+|\tau|^{2}+|\sigma|^{2}+|x|^{2}$ and $|v_{2}|^{2}+|\tau|^{2}+|\sigma|^{2}+|x|^{2}$ respectively, where we define $|q|^{2}\equiv\bar{q}q=q_{i}^{2}1_{2}$. These are, therefore, trivially satisfied in the commutative case. The off-diagonal constraints give us
\begin{equation*}
\bar{v}_{1}v_{2}+\bar{\tau}\sigma-\bar{\sigma}\tau=0
\end{equation*}
and its conjugate. These constraints may be combined as $\bar{v}_{1}v_{2}-\bar{v}_{2}v_{1}=2(\bar{\sigma}\tau-\bar{\tau}\sigma)$ and solved, in general, by \cite{Osborn:1981}
\begin{equation}\label{eq:SigmaDep}
\sigma = \frac{\tau}{4|\tau|^{2}}\left(\bar{v}_{2}v_{1}-\bar{v}_{1}v_{2}\right)+\lambda\tau
\end{equation}
for $\lambda\in\mathbb{R}$ arbitrary. The parameter $\lambda$ arises from the residual symmetry of the data, given by a transformation of $\Delta$ as
\begin{equation}\label{eq:GaugeRedund}
\Delta\to\begin{pmatrix} q & 0 \\ 0 & R \end{pmatrix} \Delta R^{-1},
\end{equation}
for $R\in\text{O}(2)$ and $q\in\text{SU}(2)$ a unit quaternion. We may choose to break this symmetry to a discrete subgroup thereof by setting $\lambda=0$.

Heuristically, then, our ADHM data contains only three independent quaternion terms and some centre of mass coordinates (suppressed inside $\tau$), which will furnish the full $16$-dimensional space.

The metric for such data has already been calculated in \cite{Allen:2013} and we will not revisit it in detail here. The salient points of the analysis are that the metric splits into two parts: a `flat' and an `interacting' part:
\begin{align}
\begin{split}
\frac{\ud s^{2}}{8\pi^{2}}&=\text{Tr}\left(\ud s_{\text{flat}}^{2}+\ud s_{\text{int}}^{2}\right) \\
&=\text{Tr}\left(\left(\ud v_{1}^{2}+\ud v_{2}^{2}+\ud \tau^{2}+\ud \sigma^{2}\right)-\frac{\ud k^{2}}{N_{A}}\right),
\end{split}
\end{align}
where
\begin{align*}
dk &= \bar{v}_{1}\ud v_{2}-\bar{v}_{2}\ud v_{1}+2\left(\bar{\tau}\ud\sigma-\bar{\sigma}\ud\tau\right), \\
N_{A} &= |v_{1}|^{2}+|v_{2}|^{2}+4\left(|\tau|^{2}+|\sigma|^{2}\right)
\end{align*}
and $\ud q^{2}\equiv \ud q\cdot \ud q=\tfrac{1}{2}\text{Tr}(\ud\bar{q}\ud q)$. This may be further simplified by explicitly writing $\sigma=\sigma(v_{1},\,v_{2},\,\tau)$ and application of a series of quaternion trace identities. Unfortunately, the noncommutative case is not so clear.

With this commutative data, one may consider scattering. Unlike in the single instanton case (where geodesic motion can avoid the singularity at zero-size by starting with non-zero angular momentum) the interactions between the two instantons can, and do, allow one instanton to shrink to zero-size in finite time. In the single instanton case, the singularity can be smoothed out by considering noncommutativity on the space, and the resulting moduli space is Eguchi-Hanson \cite{TongLee:2001}. We wish to achieve the same smoothing in the two instanton case.

\subsection{The noncommutative deformation}
Given the above, it is natural to wonder if one could deform the commutative data to encompass the effect of the noncommutativity. This is reinforced by various expected limits of the noncommutative metric: the singularity at $v_{1},\,v_{2},\,\tau\rightarrow0$ should be resolved; it should reduce smoothly to the singular, commutative, metric as we reduce the noncommutativity parameter to zero; and in the limit of large separation (that is, in the zero interaction limit) the metric should reduce to two distinct single noncommutative instanton Eguchi-Hanson metrics. Given these considerations, we deform the commutative data as follows.

To temporarily avoid confusion with the `vanilla' data above, we begin by writing the unconstrained ADHM data in the form
\begin{align*}
L&=\left(\begin{matrix} w_{1} & w_{2} \end{matrix}\right) \\
M&=\left(\begin{matrix} t & s \\ s & -t \end{matrix}\right).
\end{align*}
As remarked in \secref{subsec:ADHM}, the diagonal terms in the ADHM constraint $\Delta^{\dagger}\Delta\propto 1_{2}$ will no longer be automatically proportional to the identity but instead receive a term from $\bar{x}x$. The off-diagonal terms, having no $x$-dependence, remain the same. Hence we may still express $s$ in a similar form to the expression for $\sigma$ in \eqref{eq:SigmaDep}:
\begin{equation}\label{eq:SDep}
s=\frac{t}{4t^{\dagger}{t}}\left(w_{2}^{\dagger}w_{1}-w_{1}^{\dagger}w_{2}\right).
\end{equation}
We use the $^{\dagger}$ notation to reinforce that the entries in $\Delta$ need no longer be quaternionic.

The solution of these new constraints now results in a choice of how to perturb the quaternion parts of the commutative data $\Delta_{\text{comm}}$. The most expedient choice is to retain the quaternionic nature of $t$, which we will return to labelling as $\tau$, and absorb the noncommutativity into the $w_{a}$ as follows:
\begin{align}
\begin{split}
w_{a}&=v_{a}M_{a}, \\ 
M_{a}&=\frac{1}{\sqrt{|v_{a}|^{2}}}\left(\begin{matrix} \sqrt{|v_{a}|^{2}+\alpha\zeta} & 0 \\ 0 & \sqrt{|v_{a}|^{2}-\alpha\zeta} \end{matrix}\right),
\end{split}
\end{align}
where $\alpha=\alpha(\tau,\,v_{1},\,v_{2})$ is some function of the commutative parameters to be determined. One notes that the expression for $s$ is also no longer quaternionic, due to its form in \eqref{eq:SDep}. The constraint on $\alpha$ is given by requiring that the non-identity proportional parts of the nonquaternionic data,
\begin{equation}
\left(w_{a}^{\dagger}w_{a}+s^{\dagger}s\right)-\left(\begin{matrix} 2\zeta & 0 \\ 0 & -2\zeta \end{matrix}\right)\propto 1_{2}
\end{equation}
for $a=1,\,2$ and the solution, while non-trivial, is given by
\begin{equation}\label{eq:aForm}
\alpha=\frac{32|\tau|^{2}|v_{1}|^{2}|v_{2}|^{2}}{16|\tau|^{2}|v_{1}|^{2}|v_{2}|^{2}+|\bar{v}_{2}v_{1}-\bar{v}_{1}v_{2}|^{2}\left( |v_{1}|^{2}+|v_{2}|^{2}\right)}.
\end{equation}
It is clear at this stage why the calculation of the noncommutative metric is so much more computationally expensive than that of the commutative case. Even something as simple as the `flat' $\ud w_{a}^{2}$ is a non-trivial multi-term expansion of all of the moduli space parameters. In practice, however, we can avoid some of the complications inherent in the noncommutative metric by treating $\alpha$ as a parameter in its own right and deriving a geodesic equation for $\alpha$, containing no genuine dynamical content, whose satisfaction must be guaranteed. For later reference, this corresponds in the numerical derivation of results to introducing an additional `free' coordinate in the moduli space, along with an additional constraint in the form of a geodesic equation for $\alpha$.

Even with this simplification, calculating the metric for noncommutative instantons is not easy. Consider first the `flat' term $\ud w_{1}^{2}$. The derivative is given by
\begin{equation*}
\ud w_{1}=\ud v_{1}M_{1}+v_{1}\frac{\bar{v}_{1}\cdot\ud v_{1}}{|v_{1}|^{2}}(M^{-1}_{1}-M_{1}).
\end{equation*}
Even in the free sector of the metric, we obtain additional terms proportional to $M_{a}$. These will have minimal impact for small noncommutativity or large instantons, but in the regime where $\zeta\sim|v_{a}|$ the additional noncommutative effects will be dominant. This complexity of the deformation, even for the `free' metric terms, prevents us in all but the most simple cases from using properties of quaternion products and quaternion trace identities, as employed in \cite{Allen:2013}. A possible avenue of exploration in order to utilise such identities may be to consider the commutation relations between the quaternions and the noncommutative deformations $M_{a}$. Note that
\begin{alignat*}{2}
[M_{a},\,e_{\beta}]&=iP_{a}\epsilon_{\beta\gamma}e_{\gamma}\hspace{5mm} & \text{for $\beta,\,\gamma=1,2$,} \\
[M_{a},\,e_{i}]&=0 & \text{otherwise},
\end{alignat*}
where
\begin{equation*}
P_{a}=\sqrt{|v_{a}|^{2}+\alpha\zeta}-\sqrt{|v_{a}|^{2}-\alpha\zeta}.
\end{equation*}
We may write these commutation relations schematically as
\begin{equation}\label{eq:MComms}
[M_{a},\,e_{i}]=iP_{a}\epsilon_{ij}e_{j},
\end{equation}
where it is understood that $\epsilon_{3i}=\epsilon_{4i}=\delta_{3i}=\delta_{4i}=0$. Then we may use \eqref{eq:MComms} to collect together the factors of $M_{a}$ in the derived $s$ in \eqref{eq:SDep}. The result is
\begin{equation*}
s=\sigma M_{1}M_{2}+\frac{\tau}{4|\tau|^{2}}\frac{(\bar{v}_{2}v_{1})^{i}}{|v_{1}||v_{2}|}\left( i(|v_{2}|P_{2}M_{1}+|v_{1}|P_{1}M_{2})\epsilon_{ij}-2P_{1}P_{2}\delta_{ij}\right)e_{j}.
\end{equation*}
While this does make clearer the additional factors introduced into $s$ as a result of the noncommutativity (and indeed was used when deriving \eqref{eq:aForm}), it does not seem to provide a clear path to an explicit form for the metric without choosing a definite parametrisation.

\subsection{The moduli space and gauge transformations}
Before we select a relevant parametrisation for the metric, we first examine the effect that gauge transformations have on the derivation of the noncommutative metric for two instantons. Recall that the noncommutative ADHM data is defined up to some U$(2)$ gauge equivalence, described by \eqref{eq:GaugeRedund}. This `redundancy' corresponds to local gauge transformations of the data, and as such these transformations must be quotiented out in order to uniquely describe each point of the induced moduli space in terms of ADHM data\footnote{Any global (large) gauge transformations correspond, in the D-brane picture, to the SU$(2)$ flavour symmetry, and we implicitly include those in the ADHM parameters, $z^{r}$.}. The gauge fixing condition that removes these redundant U$(2)$ transformations is tantamount to finding the unique, time-dependent, solution to the ADHM data that satisfies the Gauss' law constraint
\begin{equation*}
D_{i}F_{i0}=0.
\end{equation*}
One can achieve this by an explicit recourse to zero modes of the data (see, for example, \cite{TongTASI:2005}). We instead follow the method of \cite{Osborn:1981}, where the zero-mode requirement degenerates to an algebraic constraint on the metric data.

To begin, we consider the metric derivation presented in \cite{Allen:2013}. We may write the metric in terms of the ADHM data as
\begin{equation}\label{eq:NaiveMetric}
g_{rs}=2\pi^{2}\text{Tr}\left(\partial_{r}a^{\dagger}(1+P_{\infty})\partial_{s}a-\left(a^{\dagger}\partial_{r}a-(a^{\dagger}\partial_{r}a)^{\text{T}}\right)\delta_{s}R\right),
\end{equation}
where $P_{\infty}$ is the projector at infinity, given in our case by $\text{diag}(1,0,0)$, and the variation $\delta R$, where $R$ is the gauge transformation in \eqref{eq:GaugeRedund}, is determined by the symmetry of the theory and the derived `zero-mode' constraint
\begin{equation}\label{eq:RConstraint}
a^{\dagger}\delta a-(a^{\dagger}\delta a)^{\text{T}}=a^{\dagger}b\delta R b^{\dagger}a-b^{\dagger}a\delta R a^{\dagger}b+\mu^{-1}\delta R +\delta R \mu^{-1}.
\end{equation}
We now consider the deformation of each term under the introduction of noncommutativity. The redundancy \eqref{eq:GaugeRedund} now requires $q\in\text{U}(2)$, rather than SU$(2)$. The `flat' terms possess no redundancy, and need no modification, under the SU$(2)$ piece of the U$(2)$, as in the commutative case, but there is an isometry corresponding to the additional U$(1)$ factor that needs to be gauged away. Generically, we have a transformation
\begin{equation*}
w_{a}\to w_{a}e^{i \xi}\;,\;\ud w_{a}\to(\ud w_{a}+i\ud\xi w_{a})e^{i\xi},
\end{equation*}
for $\xi\in\mathbb{R}$. In computing $\ud w^{\dagger}_{a} \ud w_{a}$, we must identify the conjugate momentum, $p_{\xi}$, associated to this isometry and set it to zero (this method was applied in \cite{Bak:2013} to obtain the metric of a single noncommutative instanton: one may instead define a covariant derivative acting on the ADHM data and define $\ud s^{2}=Dz^{r} D\bar{z}^{r}$). For arbitrary data $w_{a}$, after completing the square we obtain
\begin{equation}\label{eq:U1FlatDef}
\ud w_{a}^{2}=\ud\bar{w}_{a}\ud w_{a}+\vert w_{a}\vert^{2}\left(\ud \xi+\frac{\kappa}{2\vert w_{a}\vert^{2}}\right)^{2}-\frac{\kappa^{2}}{4\vert w_{a}\vert^{2}},
\end{equation}
where $\kappa=\ud\bar{w}_{a} w_{a}-\bar{w}_{a}\ud w_{a}$. The second term is equivalent to $\vert w_{a}\vert^{2} p_{\xi}^{2}$, and so must vanish. The additional U$(1)$ factor has nevertheless induced an additional factor in the flat instanton pieces. We note, at this stage, that in the limit of large separation, only the flat part of the metric contributes and $s$ vanishes. We then find that an explicit parameterisation of the $w_{a}$,
\begin{equation*}
w_{a}=\begin{pmatrix} \sqrt{\rho_{i}^{2}+\alpha\zeta}u_{a1} & -\sqrt{\rho_{i}^{2}-\alpha\zeta}\bar{u}_{a2} \\ \sqrt{\rho_{i}^{2}+\alpha\zeta}u_{a2} & \sqrt{\rho_{i}^{2}-\alpha\zeta}\bar{u}_{a1} \end{pmatrix},
\end{equation*}
for $u_{a1}=\cos\theta_{a} e^{i(\psi_{a}+\phi_{a})}$ and $u_{2a}=\sin\theta_{a} e^{i(\psi_{a}-\phi_{a})}$, results in two copies of the Eguchi-Hanson metric using the result in \eqref{eq:U1FlatDef}, as expected. In the commutative case, the expression $\kappa$ vanishes in the final metric due to the vanishing of the deformation and the presence of the trace in the metric calculation.

In the `interacting' part, the redundancy symmetries in \eqref{eq:GaugeRedund} to be parametrised by $\delta R$ now lie in U$(2)$ rather than O$(2)$, as explained in \cite{Osborn:1981}. The constraint \eqref{eq:RConstraint} is then modified accordingly. In the commutative case, the multiplicative factors around $\delta R$ were proportional to the identity, and therefore $\delta R\propto a^{\dagger}\delta a-(a^{\dagger}\delta a)^{\text{T}}$ naturally followed. In the noncommutative case, this no longer occurs. A solution is still obtainable, however: one may use the explicit $w_{a}$ and $\tau$ dependence of the data $a$ and $b$ to (anti-)commute them through $\delta R$ and explicitly multiply by the inverse of the matrix multiplicative factor. For the sake of completeness, symbolically we have
\begin{equation*}
\delta R=\left(w_{1}^{\dagger}\ud w_{2}-w_{2}^{\dagger}\ud w_{1}+2(\bar{\tau}\ud s-s^{\dagger}\ud\tau)\right)\left(w_{1}^{\dagger}w_{1}+w_{2}^{\dagger}w_{2}+ 4(\bar{\tau}\tau+s^{\dagger}s)\right)^{-1},
\end{equation*}
and so the interacting part of the metric follows trivially:
\begin{equation*}
\ud s^{2}_{\text{int}}=-\text{Tr}\left(\left(w_{1}^{\dagger}\ud w_{2}-w_{2}^{\dagger}\ud w_{1}+2(\bar{\tau}\ud s-s^{\dagger}\ud\tau)\right)^{2}\left(w_{1}^{\dagger}w_{1}+w_{2}^{\dagger}w_{2}+ 4(\bar{\tau}\tau+s^{\dagger}s)\right)^{-1}\right).
\end{equation*}
It is possible, at this point, to expand $s$ in terms of $w_{a}$ and $\tau$ and calculate the inverse but the resulting expression is not illuminating. Instead, we now exploit the symmetries of the metric to obtain tractable results.

\subsection{Complexification of the moduli space}
The noncommutative framework causes a number of complications in determining a useful form of the metric. Taking a generic parametrisation of $w_{1}$, $w_{2}$ and $\tau$ via, for example, Euler angles or complex matrices would be the easiest way to generate a full metric for the instantons, but this has proven to be computationally expensive. We may, instead, consider whether any valid geodesic submanifolds of the data exist that admit a sensible parametrisation and tractable metric calculation. Such a submanifold can be generated by certain fixed points of a symmetry of the metric. Consider the unexpanded form of $\ud s^{2}$:
\begin{equation}\label{eq:UnexpandedMet}
\ud s^{2}=\text{Tr}\left(\ud w_{1}^{\dagger}\ud w_{1}+\ud w_{2}^{\dagger}\ud w_{2}+\ud\bar{\tau}\ud\tau+\ud s^{\dagger}\ud s-N_{A}^{-1}\ud k^{2}\right),
\end{equation}
where $N_{A}$ is the multiplicative factor defined in \cite{Allen:2013}. The key symmetry that we wish to consider is conjugation of the data by a unit quaternion, $p$:
\begin{equation*}
w_{1}\rightarrow pw_{1}\bar{p},\hspace{5mm}w_{2}\rightarrow pw_{2}\bar{p},\hspace{5mm}\tau\rightarrow p\tau\bar{p}.
\end{equation*}
In the commutative picture, the invariance of the metric under such a transformation was guaranteed as the corresponding transformation rule for $\sigma$, that is $\sigma\rightarrow p\sigma\bar{p}$, is naturally respected. It is not as simple in the noncommutative case, due to the commutation relations \eqref{eq:MComms}. In order to apply the same analysis, we may only consider conjugation symmetries whose direction commutes with the direction of the noncommutativity. Clearly, then, this symmetry is valid only for $p=e_{3}$ in the noncommutative picture; the choice of noncommutativity has removed some of the underlying symmetries of the space, as would be anticipated. Our valid geodesic submanifold, then, is composed of $\tau,\,v_{1},\,v_{2}\in\text{Span}\{ e_{3},\,1_{2}\}$. Note that this complexification is in agreement with the arguments put forward in \cite{Hanany:2003}, where the $e_{3}\text{-}e_{4}$ plane is chosen in order to break the correct subgroup of the ADHM symmetries (we will examine this in more detail in \secref{subsec:Vortex}).

We thus consider an explicit complex parametrisation of the form
\begin{align*}
v_{a}&=\rho_{a}(\cos\theta_{a}1_{2}+\sin\theta_{a}e_{3}), \\
\tau&=\omega(\cos\chi 1_{2}+\sin\chi e_{3}).
\end{align*}
Due to the commuting nature of the deformation in this submanifold, then, we obtain
\begin{equation*}
s=\sigma M_{1}M_{2}=M_{1}M_{2}\sigma,
\end{equation*}
and in this parametrisation the noncommutative deformation function $\alpha$ takes on a simpler form:
\begin{equation*}
\alpha=\frac{8\omega^{2}}{4\omega^{2}+\sin\phi(\rho_{1}^{2}+\rho_{2}^{2})},
\end{equation*}
where we now define $\phi\equiv\theta_{1}-\theta_{2}$ to be the relative gauge angle on the moduli space. We also define $\Theta\equiv\theta_{1}+\theta_{2}$, corresponding to the total gauge angle.

It is now possible to calculate the metric on this $6$-dimensional submanifold. Defining, for convenience, the following quantities:
\begin{align*}
\rho^{2}_{i\pm}&\equiv\rho_{i}^{2}\pm \alpha\zeta, \\
P_{i}&\equiv\rho_{i}^{4}-\alpha^{2}\zeta^{2}, \\
\Omega_{\pm}&\equiv\rho_{1}^{2}\rho_{2}^{2}\pm \alpha^{2}\zeta^{2},\\
N_{\pm}&\equiv4\omega^{2}+\rho_{1}^{2}+\rho_{2}^{2}\pm2\alpha\zeta+\frac{1}{\omega^{2}}\rho_{1\pm}\rho_{2\pm}\sin^{2}\phi,
\end{align*}
we find the flat part to be
\begin{align*}
\ud s_{\text{flat}}^{2}=&\frac{1}{P_{1}}\left(\rho_{1}^{4}+\frac{\rho_{1}^{2}\Omega_{-}\sin^{2}\phi}{4\omega^{2}}\right)\ud \rho_{1}^{2}+\frac{1}{P_{2}}\left(\rho_{2}^{4}+\frac{\rho_{2}^{2}\Omega_{-}\sin^{2}\phi}{4\omega^{2}}\right)\ud \rho_{2}^{2} \\
&+(\ud \omega^{2}+\omega^{2}\ud \chi^{2})\left(1+\frac{\Omega_{+}\sin^{2}\phi}{4\omega^{4}}\right)+\frac{1}{4}(\rho_{1}^{2}+\rho_{2}^{2}-\frac{1}{2}\alpha^{2}\zeta^{2})(\ud\Theta^{2}+\ud\phi^{2}) \\
&+\frac{1}{2}(\rho_{1}^{2}-\rho_{2}^{2})\ud\Theta\ud\phi+\frac{\Omega_{+}\cos^{2}\phi}{4\omega^{2}}\ud \phi^{2}-\frac{\Omega_{+}\sin2\phi}{4\omega^{4}}\omega\ud\omega\ud\phi \\
&+\frac{\rho_{1}\rho_{2}\sin^{2}\phi}{2\omega^{4}}\left(\omega^{2}\ud\rho_{1}\ud\rho_{2}-\omega\ud\omega(\rho_{1}\ud\rho_{2}+\rho_{2}\ud\rho_{1})\right) \\
&+\frac{\rho_{1}\rho_{2}\sin2\phi}{4\omega^{2}}(\rho_{2}\ud\rho_{1}-\rho_{1}\ud\rho_{2})\ud\phi \\
&+\alpha\ud \alpha\zeta^{2}\left(\frac{\rho_{1}\ud\rho_{1}}{P_{1}}\left(\frac{(\rho_{1}^{2}-\rho_{2}^{2})\sin^{2}\phi}{4\omega^{2}}-1\right)+\frac{\rho_{2}\ud\rho_{2}}{P_{2}}\left(\frac{(\rho_{2}^{2}-\rho_{1}^{2})\sin^{2}\phi}{4\omega^{2}}-1\right)\right. \\
&\left.-\frac{1}{4\omega^{2}}(2\omega\ud\omega\sin^{2}\phi-\omega^{2}\sin2\phi\ud\phi)\right)\\
&+\frac{\ud \alpha^{2} \zeta^{2}\sin^{2}\phi}{16\omega^{2}P_{1}P_{2}}\left(\Omega_{-}(\rho_{1}^{2}+\rho_{2}^{2})-2\alpha^{2}\zeta^{2}(\rho_{1}^{4}-2\alpha^{2}\zeta^{2}+\rho_{2}^{4})+4\zeta^{2}\omega^{2}\Omega_{-}(\rho_{1}^{2}+\rho_{2}^{2})\right)
\end{align*}
and the interacting part, similarly, is
\begin{align*}
\ud s_{\text{int}}^{2}=&\frac{\left(\cos\phi\left(\rho_{1-}\ud\rho_{2-}-\rho_{2-}\ud\rho_{1-}\right)-2\rho_{1-}\rho_{2-}\sin\phi(\ud\Theta-2\ud\chi)\right)^{2}}{8\rho_{1-}\rho_{2-}N_{-}} \\
&+\frac{\left(\cos\phi\left(\rho_{1+}\ud\rho_{2+}-\rho_{2+}\ud\rho_{1+}\right)-2\rho_{1+}\rho_{2+}\sin\phi(\ud\Theta-2\ud\chi)\right)^{2}}{8\rho_{1+}\rho_{2+}N_{+}}.
\end{align*}
The form of the metric is perhaps not particularly simple, but one can verify the anticipated properties. In the limit of $\zeta\rightarrow0$, we see that $\Omega_{\pm}\rightarrow\rho_{1}^{2}\rho_{2}^{2}$, $P_{i}\rightarrow\rho_{i}^{4}$, $\rho_{i\pm}\rightarrow\rho_{i}^{2}$ and so $N_{\pm}\rightarrow N_{A}$, where $N_{A}$ is the multiplicative factor defined in \cite{Allen:2013}. With the vanishing of the final three lines in $\ud s_{\text{flat}}$, it is then easy to see that one recovers the commutative metric of two instantons in this limit. 

We may also verify the expected result at the large separation limit: as $\omega$ becomes large, the interacting term is subleading and $\alpha\to2\Rightarrow\ud\alpha\to0$. Ignoring the flat space $\ud\omega^{2}+\omega^{2}\ud\chi^{2}$ term, we obtain
\begin{equation*}
\ud s_{sep}^{2}=\frac{\ud\rho_{1}^{2}}{1-4\zeta^{2}/\rho_{1}^{4}}+\left(1-\frac{4\zeta^{2}}{\rho_{1}^{4}}\right)\rho_{1}^{2}\ud\theta_{1}^{2}+\frac{\ud\rho_{2}^{2}}{1-4\zeta^{2}/\rho_{2}^{4}}+\left(1-\frac{4\zeta^{2}}{\rho_{2}^{4}}\right)\rho_{2}^{2}\ud\theta_{2}^{2}.
\end{equation*}
This is two copies of the Eguchi-Hanson metric restricted to the complex subspace, which was demonstrated to be the metric of a single instanton in U($N$) gauge groups \cite{Bak:2013, Hanany:2003}.

Finally, before examining the symmetries of the metric in more detail, we note that the noncommutative metric still permits the Killing vectors $\partial_{\Theta}$ and $\partial_{\chi}$. The second vector corresponds to the overall SO$(2)$ symmetry of the flat ($\omega,\,\chi$) space geometry which, under the addition of a VEV, will remain unbroken. The vector $\partial_{\Theta}$, as justified in \cite{Bak:2013}, will contribute to the potential as
\begin{equation}
V=\frac{1}{2}g_{rs}G^{r}G^{s}=\frac{v^{2}}{2}g_{\Theta\Theta},
\end{equation}
where $v$ is the strength of the potential. Hence, for later reference, we may read off the potential term for the complexified noncommutative metric:
\begin{equation}
\begin{split}
V=&\frac{1}{2}v^{2}g_{\Theta\Theta} \\
=&\frac{1}{4}v^{2}\bigg(\rho_{1}^{2}+\rho_{2}^{2}-\frac{1}{2}\alpha^{2}\zeta^{2}-4\omega^{2} \\
&+ \frac{2\omega^{2}(\rho_{1}^{2}+\rho_{2}^{2}+4\omega^{2}-2\alpha\zeta)}{N_{-}}+ \frac{2\omega^{2}(\rho_{1}^{2}+\rho_{2}^{2}+4\omega^{2}+2\alpha\zeta)}{N_{+}}\bigg).
\end{split}
\end{equation}
We note that, in the limit as $\zeta\rightarrow0$, this agrees with the $2$-instanton commutative complexified potential given in \cite{Allen:2013}, and in the single instanton limit we obtain agreement with the complexified version of the U($1$) potential obtained in \cite{Bak:2013}. We may similarly derive the angular momentum, $L$, of the instantons, given by $g_{\Theta i}\dot{z}^{i}$, which we expect to be conserved in any subsequent geodesic motion:
\begin{equation}\label{eq:AngMom}
\begin{aligned}
L&=\frac{2}{v^{2}}V\dot{\Theta}+\frac{1}{4}(\rho_{1}^{2}-\rho_{2}^{2})\dot{\phi}+\sin^{2}\phi\left(\frac{\rho_{1+}^{2}\rho_{2+}^{2}}{N_{+}}+\frac{\rho_{1-}^{2}\rho_{2-}^{2}}{N_{-}}\right)\dot{\chi} \\
&+\frac{1}{4}\sin(2\phi)\left(\left(\frac{\rho_{2+}^{2}}{N_{+}}+\frac{\rho_{2-}^{2}}{N{-}}\right)\rho_{1}\dot{\rho}_{1}+\left(\frac{\rho_{1+}^{2}}{N_{+}}+\frac{\rho_{1-}^{2}}{N{-}}\right)\rho_{2}\dot{\rho}_{2}\right).
\end{aligned}
\end{equation}
We will explicitly verify in the following that this is a conserved quantity.

\subsection{Symmetries of the noncommutative metric}\label{sec:Syms}
We conclude this section with a brief analysis of the symmetries of the noncommutative moduli space. The solution for $s$ in the ADHM constraints allowed some freedom over a choice of constant $\tau$ term; explicitly we found
\begin{equation}
s=\frac{\tau}{4|\tau |^{2}}(w_{2}^{\dagger}w_{1}-w_{1}^{\dagger}w_{2})+\lambda \tau,
\end{equation}
for $\lambda\in\mathbb{C}$. A particular choice of $\lambda$ breaks the U($2$) gauge symmetry, represented by the ADHM transformation $\Delta\to Q\Delta R^{-1}$, down to a discrete subgroup. These discrete symmetries are quotiented when considering the moduli space metric: the fixed points of these symmetries will, upon quotienting, give rise to orbifold singularities in the moduli space. Indeed, in the commutative case, it can be seen that the zero-size singularity corresponds to such fixed points. We must consider the nature of such symmetries to ensure that the noncommutative moduli space is singularity-free, and the resulting manifold smooth.

The residual symmetries generated by $R$ may be considered as reflections or rotations of the ADHM data. We therefore have the following ADHM-invariant transformations of the data:
\begin{align*}
\tilde{w}_{1}&=w_{1}\cos\theta\mp w_{2}\sin\theta, \\
\tilde{w}_{2}&=w_{1}\sin\theta\pm w_{2}\cos\theta, \\
\tilde{\tau}&=(\cos^{2}\theta-\sin^{2}\theta)\tau\mp 2\cos\theta\sin\theta s, \\
\tilde{s}&=\pm(\cos^{2}\theta-\sin^{2}\theta)s+2\cos\theta\sin\theta\tau.
\end{align*}
Such transformations clearly leave the expression $\tilde{w}_{2}^{\dagger}\tilde{w}_{1}-\tilde{w}_{1}^{\dagger}\tilde{w}_{2}$ invariant. However, to leave $\lambda=0$ invariant we must have either $\cos^{2}\theta-\sin^{2}\theta=0$ or $\cos\theta\sin\theta=0$. Hence, the remaining discrete symmetries of $R$ are described as rotations or reflections of $\Delta$ with $\theta=n\pi/4$, $n=0,1,\dots,7$, namely the elements of the dihedral group $D_{4}$.

Now we consider each group of transformations in turn, and its action on the ADHM data.
\begin{itemize}
\item $\cos\theta=\pm1,\,\sin\theta=0$. These transformations preserve $\tau$ and $s$, and preserve or negate the signs of $w_{1}$ and $w_{2}$. The fixed point of the non-trivial symmetry occurs when $w_{i}=-w_{i}$, that is when $w_{i}=0$. This is the conical singularity encountered in the commutative case. Note that in the noncommutative picture, for generic $\zeta\neq0$ this fixed point no longer lies on the moduli space of instantons, as the noncommutative parameter bounds the instanton size from below as $\vert w_{i}\vert\ge\sqrt{\alpha\zeta}$\footnote{Note that as $\omega\to0$, $\alpha\to0$ and it would appear that the instantons may attain zero-size. In this limit, however, $s$ is the dominant term describing in the metric and the instanton sizes are more correctly described by $\vert w_{1}\pm w_{2}\vert^{2}/2$, which remain bounded.}. The action of this symmetry, therefore, does not give rise to a singularity under quotienting in the noncommutative picture, as anticipated.
\item $\cos\theta=0,\,\sin\theta=\pm1$. Such transformations again preserve $\tau$, swap the roles of $w_{1}$ and $w_{2}$ (potentially with a sign change), and may negate $s$. This corresponds to the indistinguishability of the two instantons on the moduli space. We may reinterpret this as a simple invariance under the relabelling of instantons $1\leftrightarrow2$, and obtain the previous case. The fixed points of these symmetries are, for this reason, the same zero-size instanton points as above, and may be safely ignored for the same reasons.
\item $\cos\theta=\pm\frac{1}{\sqrt{2}},\,\sin\theta=\pm\frac{1}{\sqrt{2}}$. This is equivalent to swapping $\tau$ and $s$, and redefining the $w_{i}$ as some linear combination of each other. The only fixed point of this symmetry is the `trivial' fixed point, $w_{1}=w_{2}=\tau=s=0$. As we will see in \secref{subsec:NCInstScat}, this fixed point has a geometric interpretation on the moduli space, and the `singularity' obtained has no effect on the smoothness of the underlying metric.
\end{itemize}

We may now justify the claim that noncommutativity `smooths out' the moduli space: the orbifold singularities present as a result of quotienting global gauge transformations of the ADHM data no longer appear in the noncommutative moduli space due to the new, $\zeta$-dependent, form of the $w_{i}$. This is exactly what one expects \cite{Kazuyuki:2001}. From the D$4$-D$0$ perspective, a commutative solution describes D$0$s dissolved in D$4$s; the ``small instanton'' singularities arise from the transition between the (dissolved) Higgs branch, describing Yang-Mills theory, and (separated) Coulomb branches of the D-brane theory. In the noncommutative framework, the Coulomb branch is frozen out of the worldvolume field theory, and the $\zeta\neq0$ theory allows one to describe both dissolved and separated D$0$ branes without passing through the so-called `small instanton' singularity.

This concludes the derivation and analysis of the noncommutative instanton moduli space. Via a deformation of the ADHM data, solutions to the noncommutative instanton field theory can be generated, and shown to behave as expected. While it has not been possible to find a concise, explicit form for the full $16$-dimensional metric for $2$ instantons, nevertheless a geodesic submanifold of the metric still exists and one may reliably consider the evolution and behaviour of instantons on this reduced, $6$-dimensional, space. The zero-size singularity is no longer a feature of this moduli space, achieving correspondence with the overarching D-brane picture. We may now turn to more interesting aspects of this instanton solution: evolution and scattering.

\section{Noncommutative instanton dynamics}\label{sec:Scattering}
In this section, we examine the geodesic motion of two noncommutative instantons on the induced moduli space. While the metric, and induced geodesic equations, on the complexified moduli space obtained in the previous section do not admit analytic solutions in all but the simplest considerations, a numerical approach may be taken to simulate scattering, orbiting and general behaviour of the two instantons. We first consider the case where $\langle\phi\rangle=0$ before looking at the dyonic extension to the moduli space in \secref{sec:Dyonic}. The noncommutative framework admits some surprising results, particularly with regard to stable configurations of the instantons. Finally, we briefly discuss our results in the context of the non-Abelian vortex picture and find agreement with the results described in \cite{Eto:2011}.

\subsection{Instanton scattering}\label{subsec:NCInstScat}
In order to consider the effect of scattering, we first turn to the common observation of soliton dynamics \cite{MantonSutcliffe:2004, Allen:2013} two solitons colliding head-on at small velocities often results in right-angled scattering. We note that in the metric presented in the previous section, the parameter $\omega$ admits a natural interpretation as the instanton separation. However, it is not unique in this respect. In particular, the gauge transformations that leave ADHM data $\Delta$ invariant admit an equivalent ADHM solution of the form
\begin{equation*}
\Delta^{\prime}=\begin{pmatrix} \frac{1}{\sqrt{2}}(w_{1}+w_{2}) & \frac{1}{\sqrt{2}}(w_{1}-w_{2}) \\ s & \tau \\ \tau & -s \end{pmatrix}.
\end{equation*}
Hence, we may state that $s$ has equal claim to describing the separation of the instantons. This statement is further motivated by the structure of $s$. For large $\tau$, the magnitude of $s$ is small and so in this regime the separation is adequately described by the parameter $\omega$. Conversely, for small $\tau$ it is the $s$ term that will dominate. In the case where the two parameters are of similar size, neither interpretation truly holds. More formally, the separation of the instantons is given by the eigenvalues of the lower block, $M$, of $\Delta$:
\begin{equation}\label{eq:Separation}
\lambda_{\pm}=\pm\sqrt{\tau^{2}+s^{2}}.
\end{equation}
Note that the terms in the square root are not equivalent to $q^{\dagger}q$. We interpret these as follows. For $\tau$ large, the eigenvalues are approximately $\pm\tau$ and so we identify the configuration as that of two instantons whose centres are at $\pm \vert\tau\vert$. As $\tau$ reduces, the size of $s$ is less suppressed, until we approach the point where $\tau$ and $s$ are of equal magnitude. At this point, we note that the separation \eqref{eq:Separation} vanishes since $\tau$ and $s$ are related by an imaginary phase in the commutative case. Passing beyond this point, as we reduce $\tau$ further then $s$ becomes the dominant parameter controlling separation. Right-angled scattering arises due to this interchange between $\tau$ and $s$, coupled with the imaginary multiplicative factor which causes a phase difference of $\pi/2$ between the $\tau$-dominated and $s$-dominated regimes of parameter space. In the noncommutative picture, this is not as clear. The presence of the parameter $\alpha$ in the expression for $s$ makes the zero-eigenvalue requirement more complicated, and the results are dependent on the magnitude of $\zeta$. 

\begin{figure}[!ht]
\cincludegraphics[width=0.8\textwidth]{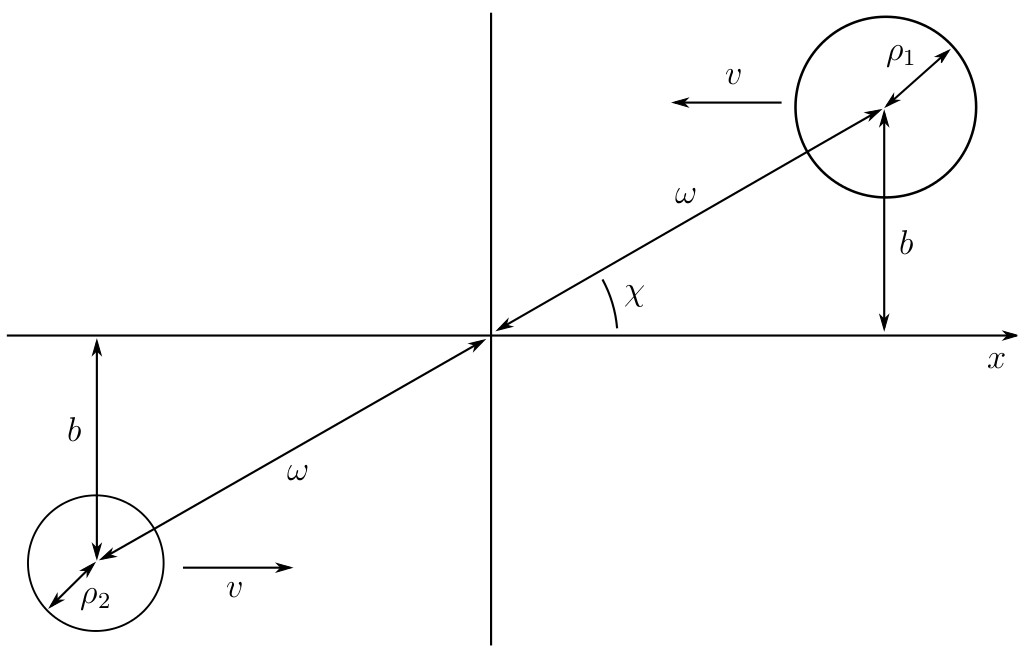}
\caption{The relevant parameter set-up for scattering simulations. The general separation of the instantons is described by $1/\sqrt{2}(\vert \tau\vert^{2}+\vert s\vert^{2})$, and this is what the ``x'' and ``y'' axes describe. In subsequent plots, where it is helpful, we plot the sizes of the instantons at regular $t$-intervals to demonstrate size evolution and instanton speed.}
\label{fig:Scat}
\end{figure}

The scattering scenario is shown in \figref{fig:Scat}. Using the complexified metric derived previously, we identify $\rho_{i}$ with the size of the $i$-th instanton. The angle $\chi$ defines the angle of incidence of scattering relative to the axis (so that an angle of $\chi=0$ represents head-on scattering) and $\omega$ the initial separation. Due to the discontinuous jump that occurs at zero separation (representing the symmetry between $w_{1}$ and $w_{2}$), a naive numerical simulation breaks down at the point of collision. We thus follow \cite{Allen:2013} and reparametrise the variables in $\tau$ as $\omega=\sqrt{x^{2}+b^{2}}$ and $\chi=\arctan(b/x)$. Then we may interpret $x$ as the initial separation along the axis and $b$ as an impact parameter. A head-on collision will occur when the impact parameter goes to $0$ but can be approximately observed for sufficiently small, non-zero, $b$.

\subsection{Head-on Collisions}

\begin{figure}[!ht]
\cincludegraphics[width=0.9\textwidth]{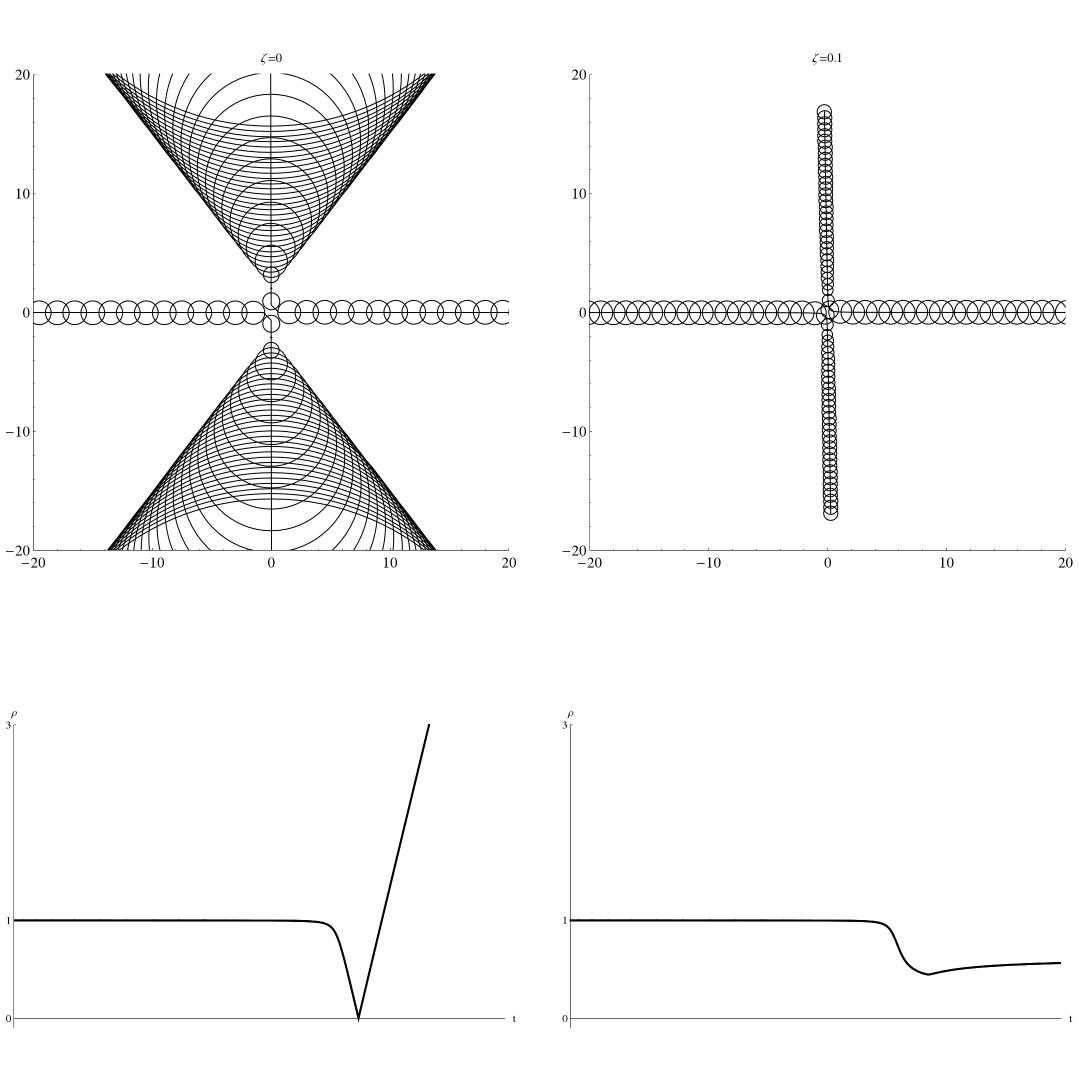}
\caption{A comparison of commutative (left) and noncommutative (right) instanton scattering and sizes for given initial conditions $\phi=\pi/2,\,b=0.001,\,x=30\text{ and }\rho_{1}=\rho_{2}=1$. Right-angled scattering is still a valid behaviour in the noncommutative case for small impact parameter. We note that, as anticipated, the instanton sizes do not vanish at the point of collision, thus avoiding the moduli space singularity attained in the commutative case.}
\label{fig:CVsNC1}
\end{figure}

\begin{figure}[!ht]
\cincludegraphics[width=0.8\textwidth]{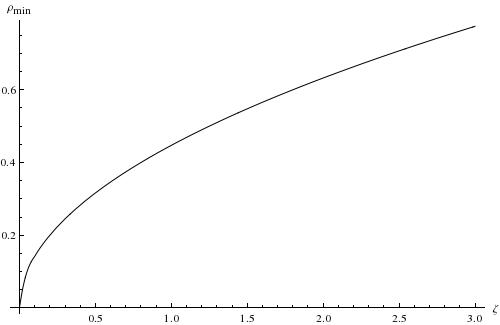}
\caption{Minimum instanton size achieved via head-on scattering with varying $\zeta$.}
\label{fig:NCSize}
\end{figure}

\begin{figure}[!ht]
\cincludegraphics[width=0.8\textwidth]{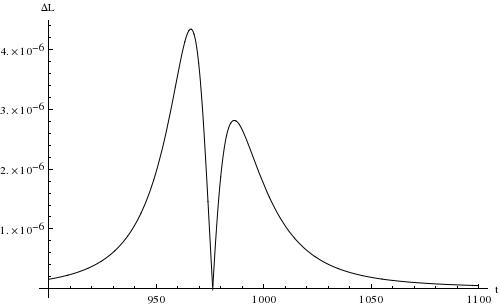}
\caption{The variation in angular momentum of the system around the point of collision. The difference between the initial angular momentum and that of the evolved configuration never exceeds $\mathcal{O}(10^{-6})$, well within numerical error. Outside of the scattering region, the difference drops to $\mathcal{O}(10^{-8})$.}
\label{fig:AngMom}
\end{figure}

We first consider the results of such a `head-on' collision in both the commutative ($\zeta=0$) and noncommutative ($\zeta=0.1$) systems, as shown in \figref{fig:CVsNC1}. The presence of right-angled scattering is perhaps heartening, as this agrees with the expected soliton behaviours. The key point, however, lies in the size plots. While in the commutative framework the instanton sizes reach the zero-size singularity, no such problem exists in the noncommutative analogue. This is as expected, since one anticipated that the noncommutativity would smooth out the singular point encountered in the commutative picture. It can be verified that the minimum of the size is attained just after collision, and with the parameters evaluated, this minimum is precisely $\sqrt{\alpha\zeta}$. This agrees with our expectations: the noncommutative deformation to the metric took the form $\rho^{2}_{i}\rho^{2}_{j}-\alpha^{2}\zeta^{2}$ for $i=1,\,2$ and $j=1,\,2$ so the singularities at $\rho_{i}=0$ are replaced by a circle around the $\rho_{i}$ parameter spaces of size $\sqrt{\alpha\zeta}$. This trend is shown in \figref{fig:NCSize}. Finally, we note that angular momentum \eqref{eq:AngMom} is conserved: the period of greatest volatility is around the point of collision. In this regime, the difference between initial angular momentum and that of the scattering configuration varies only slightly, and well within expected numerical error. The ``change'' in angular momentum is shown in \figref{fig:AngMom}. 

\begin{figure}[!ht]
\cincludegraphics[width=0.9\textwidth]{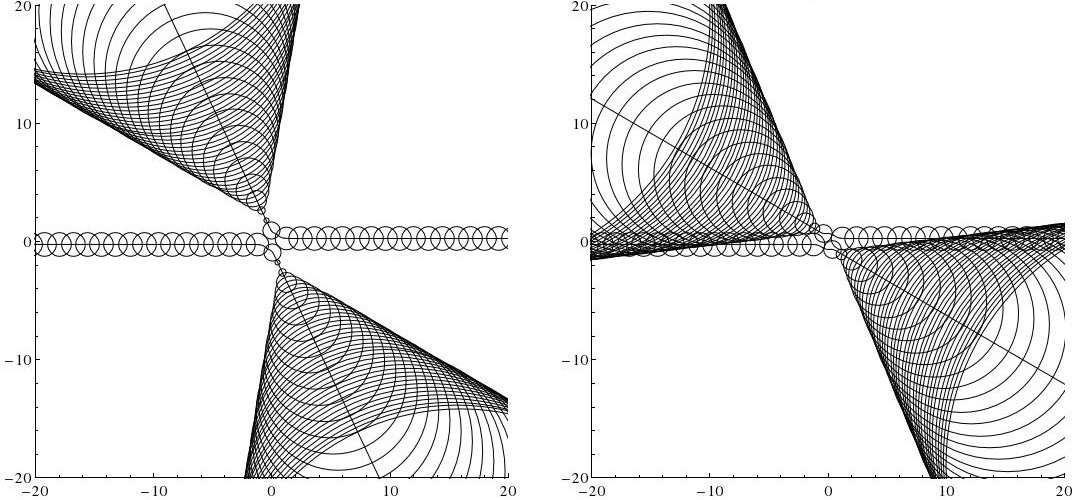}
\caption{Commutative and noncommutative scattering for $b=0.25$.}
\label{fig:CVsNC2}
\end{figure}

We may also move away from the head-on limit of scattering and consider a non-negligible impact parameter. In the commutative case, this allows the instantons to deviate away from $\pi/2$ scattering: in the noncommutative case, this effect is even more pronounced. This behaviour is shown in \figref{fig:CVsNC2}.

The above demonstrates that the `attractiveness' of the noncommutative bound state displays a large sensitivity to the value of the impact parameter, $b$. As one varies the impact away from head-on, we obtain scattering, although the presence of $\zeta\neq0$ deforms the scattering solutions away from the commutative scattering angle. This behaviour under introduction of noncommutativity appears to be a generic feature of all soliton systems which arise from reductions of noncommutative instantons: in considerations of non-Abelian vortices (where a Fayet-Iliopoulos parameter serves to couple the Abelian U($1$) non-trivially to the rest of the gauge group), this attractive behaviour is also manifest \cite{Eto:2006}. It is natural to ask whether such an attractive force on the moduli space could be interpreted as an induced potential on the space, even for the free instanton moduli space. This is a question that we will revisit in due course.

\begin{figure}[!ht]
\begin{subfigure}[t]{0.49\textwidth}
\includegraphics[width=\textwidth]{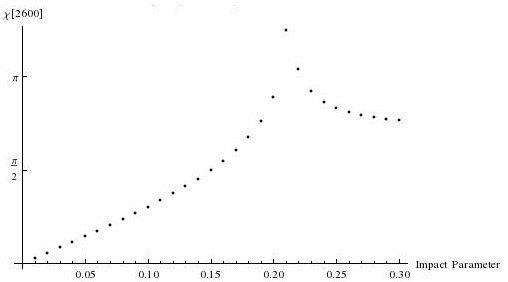}
\end{subfigure}
\begin{subfigure}[t]{0.49\textwidth}
\includegraphics[width=\textwidth]{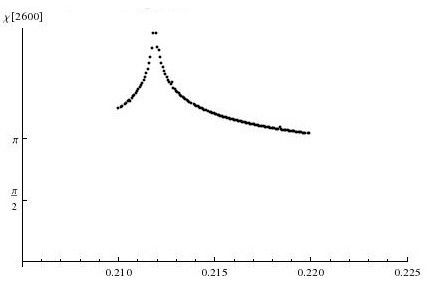}
\end{subfigure}
\caption{Scattering with varying $b$ and $\zeta=0.1$. A `critical' point in configuration space exists at $b\sim0.21$, where the instantons temporarily orbit before scattering.}
\label{fig:ChiSense}
\end{figure}

Given the modifications to the scattering behaviour under the introduction of a non-zero $\zeta$, it is instructive to examine the scattering angles obtained. The results are shown in \figref{fig:ChiSense} for equal size instantons (since this provides right-angled scattering in the commutative case) and $\zeta=0.1$. We note that as we vary the impact parameter, the scattering angle varies accordingly from standard scattering to a scattering angle greater than $\pi$. This demonstrates that, far from being the standard result, right-angled scattering is one possible outcome from the collisions of noncommutative instantons.

\begin{figure}[!ht]
\cincludegraphics[width=0.6\textwidth]{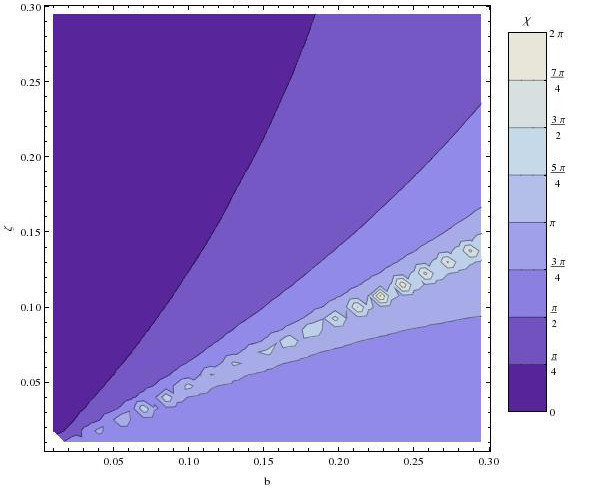}
\caption{Contour plot of final scattering angle with varying $\zeta$ and $b$. The region bounded by the countour $\chi=\pi$ contains configurations with the unstable orbit characteristics.}
\label{fig:Contour}
\end{figure}

The above results raise more questions about the behaviour of the instantons. From \figref{fig:ChiSense} one can see that there appears to be a ``critical'' tuning between $b$ and $\zeta$ which maximises the final scattering angle. Such a tuning exists for all possible values of $b$ (or equivalently, $\zeta$), as can be seen in \figref{fig:Contour}. 

\begin{figure}[!ht]
\begin{subfigure}[t]{0.49\textwidth}
\includegraphics[width=\textwidth]{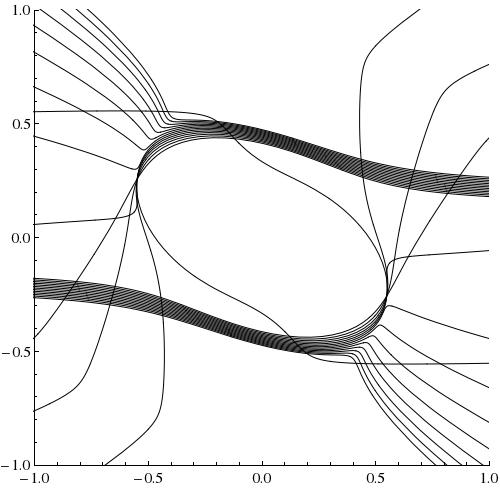}
\end{subfigure}
\begin{subfigure}[t]{0.49\textwidth}
\includegraphics[width=\textwidth]{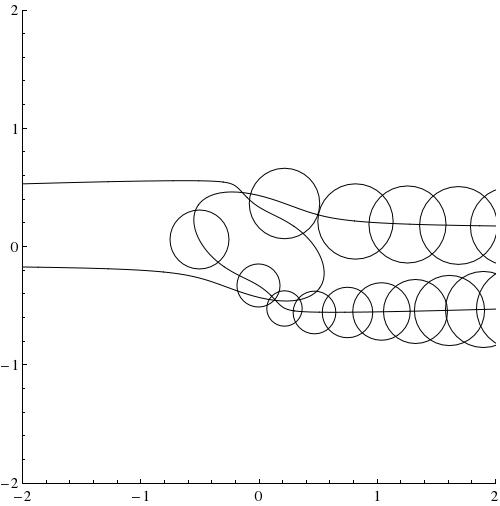}
\end{subfigure}
\caption{Collisions for a range of $b$ and $\zeta=0.1$. The configuration that maximises the scattering angle (the right figure) corresponds to a ``slingshot'', where the instantons orbit each other before returning whence they came. In the right-hand plot (where the size of one instanton has been suppressed for clarity), the right instanton approaches from above the $x$-axis with stable size and speed and leaves more slowly, but with an increasing size.}
\label{fig:UnstOrb}
\end{figure}

These results are perhaps surprising: right-angled scattering does appear, but is not the most general result for close to head-on collisions between two noncommutative instantons. In fact, it naturally arises from a consideration of the symmetries in \secref{sec:Syms} and the expression for the separation \eqref{eq:Separation}. The more involved form of $s$, coupled with the presence of the parameter $\alpha$ in the data, allows for a greater range of initial data causing the $\tau$-$s$ identification change. As a result, one can obtain scattering in a range of scenarios and scattering angles, of which right-angled is but one aspect.

\subsection{The connection to vortices}\label{subsec:Vortex}
The results gained for instantons have wider reach to other solitonic systems. The $(4+1)$-dimensional Yang-Mills theory can be dimensionally reduced in a number of ways to obtain other lower-dimensional theories. Accordingly, instantons (as solutions to bosonic Yang-Mills theory) can be dimensionally reduced to produce monopole and vortex solutions.

The vortex picture is an interesting one: the vortices are static solutions to a $(2+1)$-dimensional maximally supersymmetric $\mathcal{N}=4$ field theory. To guarantee the existence of vortex solutions, the bosonic Lagrangian of such a theory can be adapted to contain a Fayet-Iliopoulos parameter, which modifies the D-term constraints and ensures symmetry breaking of the vacuum \cite{Hanany:2003}. The introduction of such a term mediates the coupling between the SU$(N)$ gauge symmetry and the remnant U$(1)$ symmetry, in a similar vein to the instanton picture and hence the vortex solutions thus obtained can be considered to be non-Abelian \cite{Manton:2010}. The equivalence between the instanton and vortex deformations is not quite straightforward, however.

To make clear the connection, we must consider the symmetries of the instanton data \cite{Hanany:2003}. The full symmetry group of the ADHM data for U$(N)$ instantons is
\begin{equation*}
G_{\text{inst}}=\text{SO}(5)\times\text{U}(N)\times\text{SU}(2)\times\text{U}(1),
\end{equation*}
where the SO$(5)$ rotates the transverse scalars $X^{I}$, the U$(N)$ is the overall flavour symmetry (corresponding to the ADHM redundancies) and the SU$(2)\times$U$(1)$ symmetry is the unbroken parts of the worldvolume SO$(4)$ symmetry after the introduction of noncommutativity. The vortex theory arises via a symmetry breaking of a subgroup of $G_{\text{inst}}$ to leave the matter content and SUSY structure equivalent to that of the vortices. To achieve this, we weakly gauge a U$(1)$ factor inside Spin($5$). We can interpret this in a more concrete sense via the ADHM data and corresponding moduli space. The U$(1)$ gauge field is tantamount to a circle action on the moduli space, which will have a corresponding triholomorphic Killing vector $\hat{k}$. Gauging by this $S^{1}$ action leads to a potential term in the instanton Lagrangian, with mass term proportional to $\hat{k}^{2}$. Now, considering the fixed points of the circle action (equivalently, all points in the moduli space for which $\hat{k}=0$) gives us exactly the vortex moduli space. To ensure isometry between the two sets of theories, one must relate the instanton noncommutative parameter, $\zeta$, to the gauge coupling of the vortex theory; namely,
\begin{equation}\label{eq:zetae}
\zeta=\frac{\pi}{2e^{2}}.
\end{equation}

There are a number of open questions in this analysis, most of which are unfortunately beyond the scope of this work. The FI parameter in the vortex theory already guarantees the existence and smoothness of vortex solutions, unlike in the overarching instanton theory. Due to the identification between $\zeta$ and the gauge coupling of the vortex U$(1)$, descending to a theory of vortices from noncommutative instantons may, rather than resolving the moduli space, lead to singularities not present in the original theory \cite{Hanany:2003}. More work on this aspect of the analysis, including classifying such potential singularities, would be helpful.

The scope of vortex solutions, a priori, appears to be larger than those configurations that would arise from the instanton reduction. The instantons, when dimensionally reduced, provide a `critically coupled' non-abelian vortex theory and in fact, one can see from \eqref{eq:zetae} that in the commutative limit the U$(1)$ part of U$(N)$ is frozen out of the theory. However, the theory of vortices may also admit its own FI parameter as well as the U$(1)$ gauge coupling. It would appear that, as $\zeta$ is in some sense determined by the coupling $e$, that the instanton theory says nothing about the noncommutative structure of the vortex theory. It seems incongruous to assert that the vortices have additional freedom not possessed by the instantons, but a clear justification of the converse would be preferred. As it stands, we may only consider equivalence of our solutions to this critically coupled theory.

A point that naturally stems from the above discussion is related to dyonic instantons. If we choose a potential for the dyonic instantons in a direction orthogonal to the U$(1)\subset\text{Spin}(5)$, then we should anticipate some form of `dyonic' vortices to appear. The nature of such a theory is not clear, but work is being done to include a Higgs field to the vortex picture (e.g. \cite{Collie:2009}), which should find some analogue in the instanton secnario.

We do not wish to enter into a discussion of these issues. The aim here is to ensure that our solutions to the noncommutative ADHM equations agree, upon reduction, with the known behaviours of non-abelian vortices \cite{Eto:2011}. We note that, in the `free' instanton case, we have no real restriction on the choice of $\text{U}(1)\subset\text{Spin}(5)$ to gauge. We may also note that when complexifying the moduli space of instantons, we required the data to be unchanged under conjugation by the unit quaternion $e_{3}$. The data we have obtained, then, is fixed under the circle action generated by $e_{3}$ and hence viable as a starting point for the comparison with vortices.

\begin{figure}[!ht]
\begin{subfigure}[t]{0.45\textwidth}
\includegraphics[width=\textwidth]{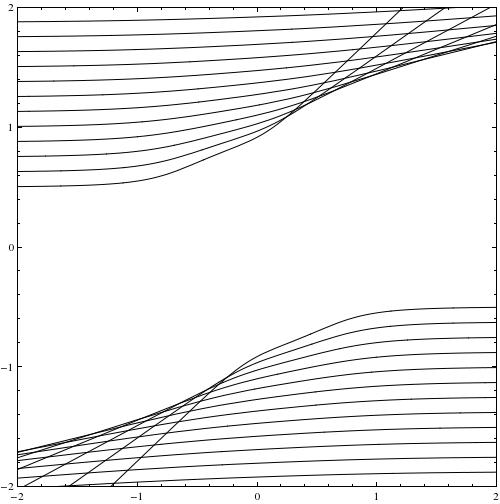}
\caption{Vortex scattering for $\phi=\pi/2$, $b=0.5,0.75,\dots$.}
\label{subfig:VortexCImpact}
\end{subfigure}
\begin{subfigure}[t]{0.45\textwidth}
\includegraphics[width=\textwidth]{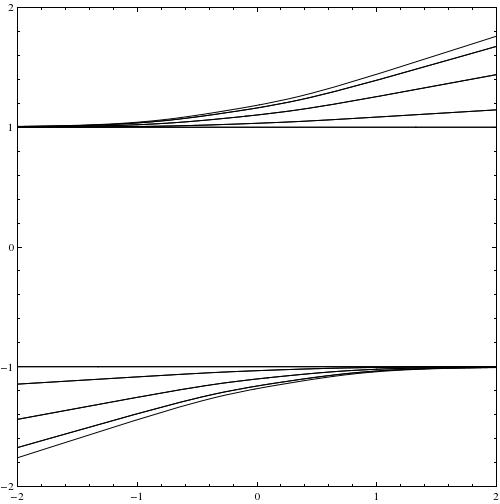}
\caption{Vortex scattering for $b=1$, $\phi=0,\pi/8,\dots$.}
\label{subfig:VortexCAngle}
\end{subfigure}
\caption{Vortices from a reduction of the instanton moduli space.}
\label{fig:Vortices}
\end{figure}

\figref{fig:Vortices} shows the results of this vortex limit, We reproduce the results gained in \cite{Eto:2011} from the instanton data and observe the expected behaviour: the non-abelianisation is shown in the different behaviours with varying gauge orientation $\phi\equiv\theta_{1}-\theta_{2}$. Of course, this is just one aspect of non-Abelian vortex scattering, but nonetheless it is encouraging to see the scattering behaviour exactly reproduced in the context of instantons.

\subsection{The attractiveness of noncommutativity}\label{subsec:NCattract}
Before moving on to consider dyonic noncommutative instantons, we may analyse the effect of noncommutativity on the free instanton picture. We noted that the presence of non-zero $\zeta$ seems to introduce an attractive effect to the normal instanton scattering that, for sufficiently high $\zeta$, overrides the normal repulsion of the two instantons. Then, before we concern ourselves with an additional potential force, we should investigate whether we may view the noncommutative effect as a genuine attractive effect. If so, then we would expect the transition to dyonic instantons to be unremarkable: the same scattering solutions will exist, but each solution will correspond to a two-parameter space spanned by the strength of the potential, $v$, and the noncommutativity.

\begin{figure}[!ht]
\cincludegraphics[width=0.9\textwidth]{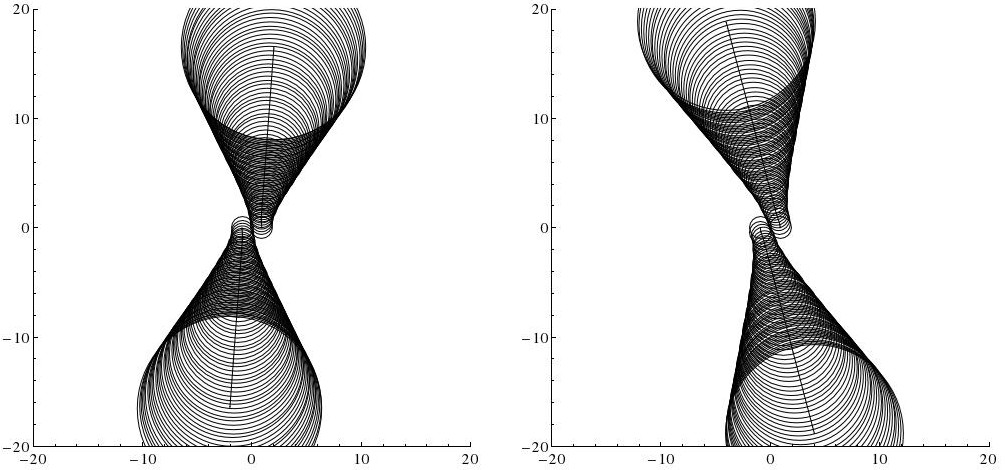}
\caption{A demonstration of the attractive effect of noncommutativity: overlapping instantons with initial motion at an angle $\pi/2$ to the $x$-axis. On the left, for $\zeta=0.05$, repulsive behaviour dominates short-scale interactions; on the right, for $\zeta=0.3$, attractive behaviour is the key feature.}
\label{fig:NCPull}
\end{figure}

The clearest possible test of this is the following. We set our instantons at a finite distance apart such that in the commutative $v=0$ case the repulsive behaviour is apparent. Sending the two initially at right angles to the line of separation, we would expect a deviation away from $\pi/2$ for a small time, until the instantons are suitably far away that repulsion is no longer a feature of the system. We may then repeat this for some appreciably large value of $\zeta$. The results are shown in \figref{fig:NCPull} for unit-size instantons and initial separation $0.9$. Crucially, the separation is chosen such that the extent of the instantons initially overlap, and so interaction effects are the dominant initial contribution to the instanton dynamics.

\begin{figure}[!ht]
\cincludegraphics[width=0.9\textwidth]{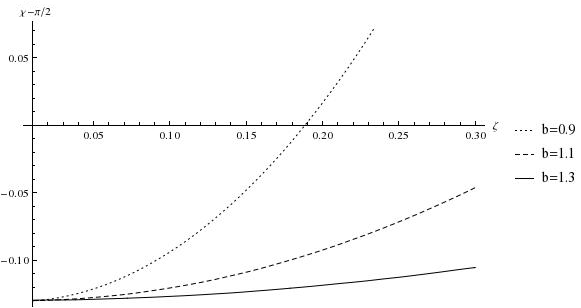}
\caption{The attractive/repulsive interface for noncommutative instantons. For $\chi-\pi/2<0$, repulsion occurs. For suitably small initial separation, one can instead obtain $\chi-\pi/2>0$ (attraction).}
\label{fig:NCAttract}
\end{figure}

On the left hand side, with $\zeta=0.05$, we observe the expected behaviour. The instantons temporarily repel, before maintaining a steady course. On the right hand side, for $\zeta=0.3$, a very different picture emerges. Far from repelling, the short-distance behaviour is attractive, before the instantons separate too far for interaction effects to dominate. If the noncommutativity is strong enough, then the tendency is for the instantons to come together rather than pull apart. \figref{fig:NCAttract} shows the changing angle of exit with varying $\zeta$ for some values of the separation, where the crossover point between repulsion and interaction becomes clear.

\figref{fig:NCAttract} also shows the sensitivity of such behaviour to the initial separation. We plot the value of the final scattering angle $\chi-\pi/2$ against $\zeta$ for a variety of impact parameters. When the plots remain below the $x$-axis, the instantons are scattering repulsively; when they cross the axis, this demonstrates the transition to attractive scattering. For large initial separation, the generic instanton repulsion is the only notable effect on the dynamics due to the subleading nature of the $\zeta$ modifications to the metric, and the crossover between repulsion and attraction is not evinced. Note that the trajectories of the plots suggests that the case $b=1.1$ will eventually cross the transition point. However, the value of $\zeta$ at which it does so is outside the valid parameter regime for the geodesic approximation and therefore cannot be considered to be a feature of the system. Nevertheless, it can be seen that the introduction of a noncommutative parameter to the moduli space can cause an attractive, rather than repulsive, effect.

We have now seen the important effects of a noncommutative parameter on instanton scattering. Far from a simple modification to scattering angle, we may observe very different behaviours. For an initially small impact parameter, we may recover the standard results of soliton scattering. However, for off-centre scattering configurations, the presence of $\zeta$ can effect an attractive force between the two instantons, greatly modifying their scattering behaviour. We now turn on an actual potential force in the metric, and consider the twin effects of the two attractions.

\section{Dyonic noncommutative instantons}\label{sec:Dyonic}
In this section, we consider the effect on the dynamics of noncommutative instantons under the addition of a potential term. The ADHM construction in \secref{sec:Construction} demonstrated that the potential term does not remain unchanged after we consider a noncommutative space. We would expect, then, that the dynamics of such instanton solutions should change accordingly.

We first recall the form of the potential term for two noncommutative U$(2)$ instantons:
\begin{align*}
V=&\frac{1}{4}v^{2}\bigg(\rho_{1}^{2}+\rho_{2}^{2}-\frac{1}{2}\alpha^{2}\zeta^{2}-4\omega^{2} \\
&+ \frac{2\omega^{2}(\rho_{1}^{2}+\rho_{2}^{2}+4\omega^{2}-2\alpha\zeta)}{N_{-}}+ \frac{2\omega^{2}(\rho_{1}^{2}+\rho_{2}^{2}+4\omega^{2}+2\alpha\zeta)}{N_{+}}\bigg),
\end{align*} 
where $N_{\pm}\equiv4\omega^{2}+\rho_{1}^{2}+\rho_{2}^{2}\pm2\alpha\zeta+\frac{1}{\omega^{2}}\rho_{1\pm}\rho_{2\pm}\sin^{2}\phi$ and $\rho_{i\pm}\equiv\sqrt{\rho_{i}^{2}\pm\alpha\zeta}$. In the extremal limit as noncommutativity becomes comparable to instanton size, that is $\alpha\zeta\sim\rho^{2}$, $N_{+}\to4\omega^{2}+4\rho^{2}+2\rho^{2}/\omega^{2}$ and $N_{-}\to 4\omega^{2}$. Then as $\omega\to0$, we see that both interacting terms become negligible and the effect of the potential term on the full dynamics is dominated by the `free' terms therein, with some marginal noncommutative modification arising from the $\alpha^{2}\zeta^{2}/2$ term. For small noncommutativity (in line with the requirements of the Manton approximation) this effect, too, is subdominant. Conversely, as previously mentioned (and studied in \cite{Allen:2013}), the commutative limit gives a similar picture: the $N_{A}$ term is subleading in the scattering limit. Hence any substantive effects of the introduction of noncommutativity are not to be found in straightforward scattering. Nevertheless, the difference between dyonic instantons and their regular counterparts may be seen in some aspects of scattering in a neighbourhood around $\omega=0$, and we may consider those. Moreover, dyonic instantons may exhibit a feature not present in the free case: it is possible to `tune' the latent repulsive force of the instantons and the attractive potential force to obtain stable orbiting solutions. We shall examine whether such solutions are an option in the noncommutative framework.

\subsection{The dyonic picture}\label{sec:NCdyonic}
Now we introduce a non-zero potential strength, $v$. The results of \secref{subsec:NCattract} were suggestive of a potential-like force on the moduli space arising from the noncommutativity. A potential of the form \cite{Peeters:2001, Allen:2013} is also useful, however, as it allows us to examine whether the slow-roll instability as $\rho\to\infty$ exists in the noncommutative case. It may also shed some light on the BPS spectra, via an analysis of the zeros of the potential \cite{Bak:2013}.

\begin{figure}[!ht]
\begin{subfigure}[t]{0.45\textwidth}
\includegraphics[width=\textwidth]{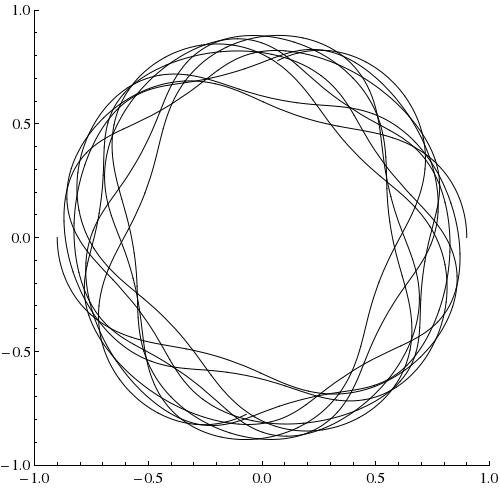}
\caption{Orbiting from perpendicular instantons, for $v=0.07$.}
\label{subfig:NCLike}
\end{subfigure}
\begin{subfigure}[t]{0.45\textwidth}
\includegraphics[width=\textwidth]{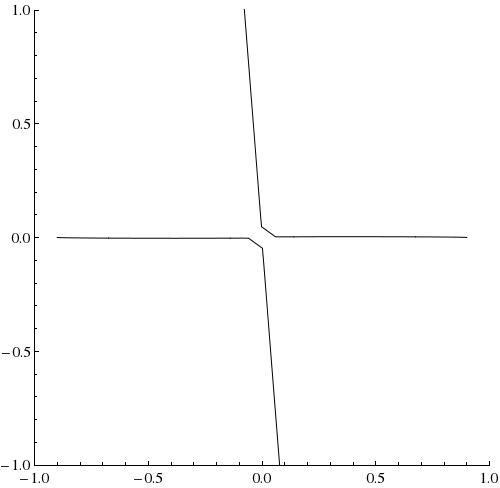}
\caption{Beyond Manton scattering: $v=10$. Objects scatter as if propelled inwards to begin with.}
\label{subfig:NotNCLike}
\end{subfigure}
\caption{The attraction options for commutative dyonic `instantons'. The instantons either attract and reside in a fixed orbit, or attract with such force that scattering occurs. No intermediate behaviour is demonstrated.}
\label{fig:PotentialTurning}
\end{figure}

The results are shown in \figref{fig:PotentialTurning}. This demonstrates quite different characteristics: for relatively low potential strength, the instantons can attract and form a stable orbit (of which we will see more shortly), with the potential force and repulsive force balanced. Even if one breaks the Manton approximation by allowing $\vert v\vert^{2}>1$, the `instanton' solutions attract so strongly that the configuration resembles that of a head-on collision. There is no configuration that envinces attractive behaviour of the form seen in the pure instanton case.

This is interesting, but perhaps not surprising. The dynamics of noncommutative instantons are resulting from purely geodesic motion: that is, any scattering effect arises due to the geometry of the moduli space. Since the key feature of the noncommutative moduli space is that the singularity at zero is smoothed out, the instantons are liable to reside in a lower energy state due to their newly allowed closeness. In the dyonic commutative picture, the singularity at zero-separation remains, and the instantons are unable to bind. Any deviation from the geodesic motion effected by the potential or a velocity will not overcome the singularity at the origin.

\begin{figure}[!ht]
\cincludegraphics[width=0.9\textwidth]{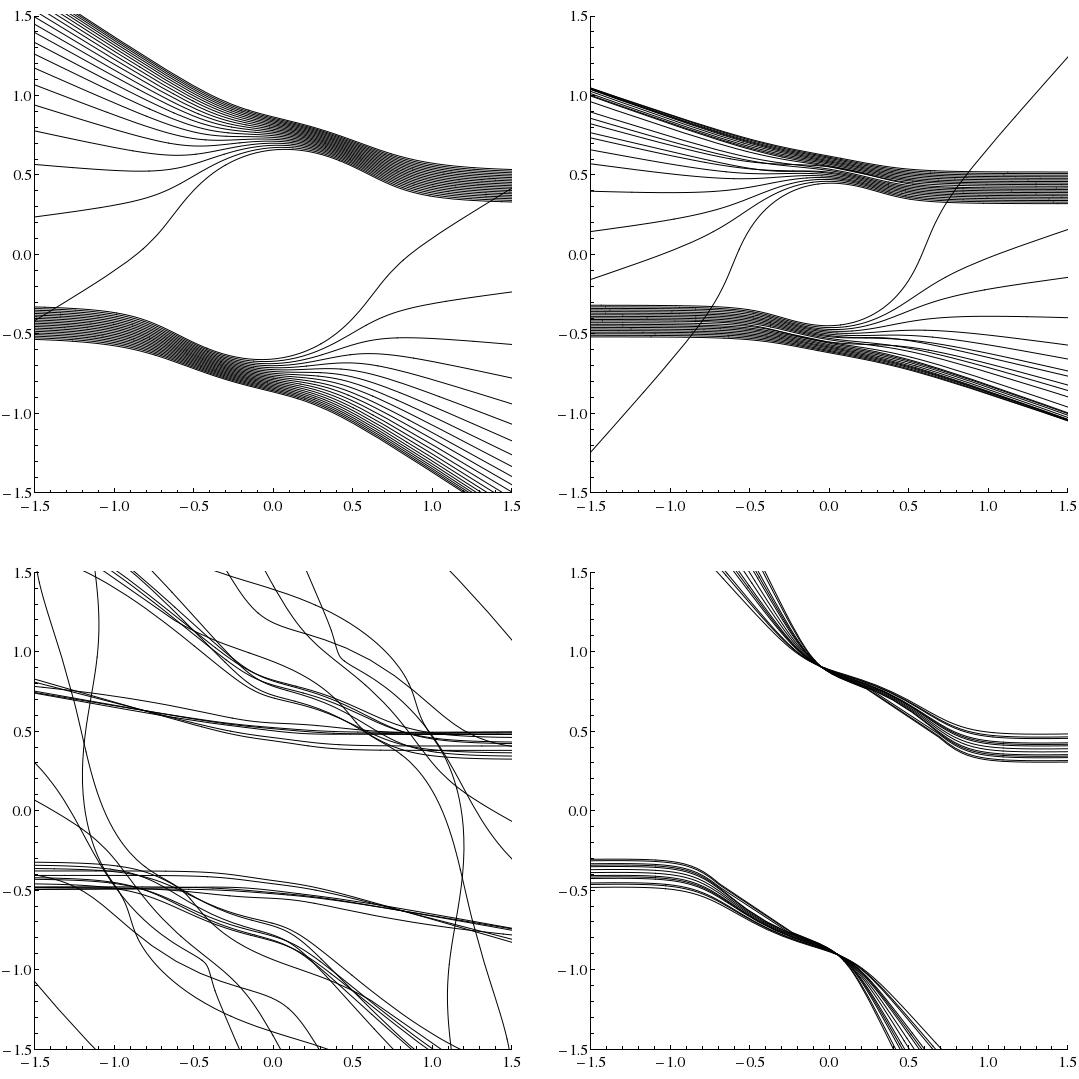}
\caption{Dyonic noncommutative scattering for $\zeta+v=0.15$. The free noncommutative result $\zeta=0.15$ is shown in the top left, followed by $\zeta=0.1,\,v=0.05$, $\zeta=0.05,\,v=0.1$ and $v=0.15$ (commutative dyonic) respectively.}
\label{fig:CombinedNCV}
\end{figure}

This aside, we consider the available solutions under the influence of both $\zeta$ and $v$. In the search for interesting results, we ignore some regions of the parameter space: the addition of an attractive potential term is not going to change the scattering behaviour for small impact parameter. Rather, we will focus on the regions of parameter space where scattering did occur in the free noncommutative picture and analyse any changes that arise in those situations. In the following, we consider a range of initial impact parameters, $0.32\le b \le 0.52$, and stipulate that the combined `strength' of the noncommutativity and potential are fixed. \figref{fig:CombinedNCV} shows the results for different partitions of $\zeta+v=0.15$, where this partition and strength are chosen in order to demonstrate the salient qualitative behaviours.

\begin{figure}[!ht]
\centering
\begin{subfigure}[t]{0.45\textwidth}
\includegraphics[width=\textwidth]{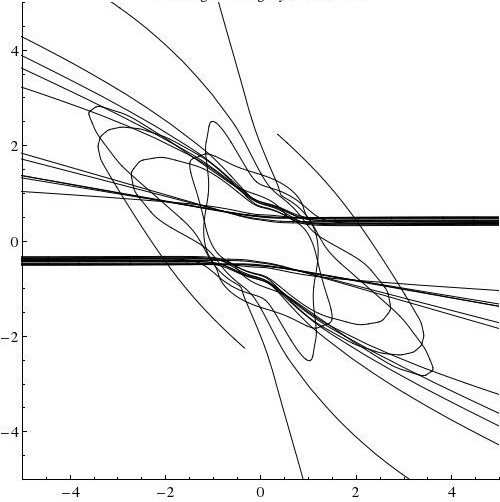}
\end{subfigure}
\begin{subfigure}[t]{0.45\textwidth}
\includegraphics[width=\textwidth]{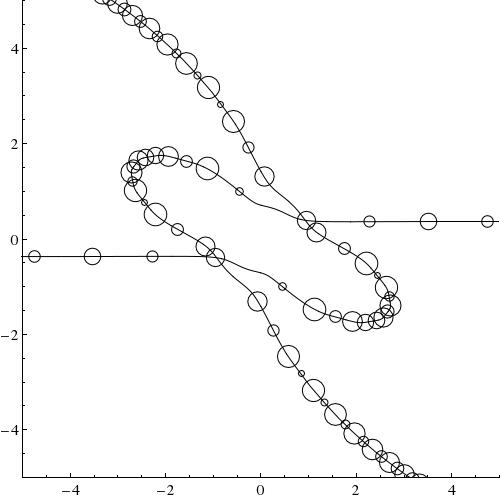}
\end{subfigure}
\caption{A zoomed out plot of the $\zeta=0.05,\,v=0.1$ configuration above, and one particular unstable orbit from the initial plot with size oscillation shown.}
\label{fig:NCUnstable}
\end{figure}

In the first case, we consider pure noncommutativity. This is a familiar result: we have a modified scattering picture. As we dial down $\zeta$ and dial up $v$, we may see very different behaviours. While $\zeta$ dominates, the pure noncommutative picture is still approximately valid; as the effect of the potential dominates, then scattering is guaranteed, albeit with the expected changes to the final scattering angle. Somewhere around the midpoint of this transition (demonstrated in \figref{fig:CombinedNCV} for $\zeta=0.05$ and $v=0.1$), the behaviour becomes more interesting. A zoomed out version of this plot is shown in \figref{fig:NCUnstable}, and shows the presence of unstable orbits even without the initial conditions chosen by \cite{Allen:2013}.

One point to make with regards to these results is that the qualitative difference between configurations with similar initial conditions can be considerable. The moduli space is incredibly sensitive to any adjustments to impact parameter and potential strength, in particular. This is not surprising: given the respective instabilities inherent in both the dyonic commutative and free noncommutative instanton configurations, a combination thereof allows for a greater range of unstable dynamical systems in the $\zeta\text{-}v$ parameter space.

\subsection{Dyonic Orbits}
Despite the observed instability of certain scattering scenarios as in \secref{sec:NCdyonic}, and the atypical behaviour of some `non-scattering' situations as in \secref{subsec:NCattract}, we may examine whether the stable orbits known to exist in the commutative picture remain in the noncommutative analogue. Such orbits existed at a point of equilibrium between the attractive and repulsive forces of the potential and the instanton effect, respectively. Given our previous considerations, it is not clear whether such a situation may be replicated for noncommutative instantons.

One key point in the search for such systems is that of longevity: the presence of a non-zero $\zeta$ has introduced the possibility of attraction and scattering for previously normal scattering scenarios, if $\zeta$ is large enough. This option is still possible if we start with a stable orbit and turn on noncommutativity, but the time taken to demonstrate the behaviour may be much longer. With this in mind, all numerical simulations run to investigate the possibility of orbits have been run for around $5$ times longer than those in \cite{Allen:2013} to rule out eventual scattering. We again consider the interplay between the noncommutative parameter $\zeta$ and the strength of the potential $v$.

\begin{figure}[!ht]
\cincludegraphics[width=0.9\textwidth]{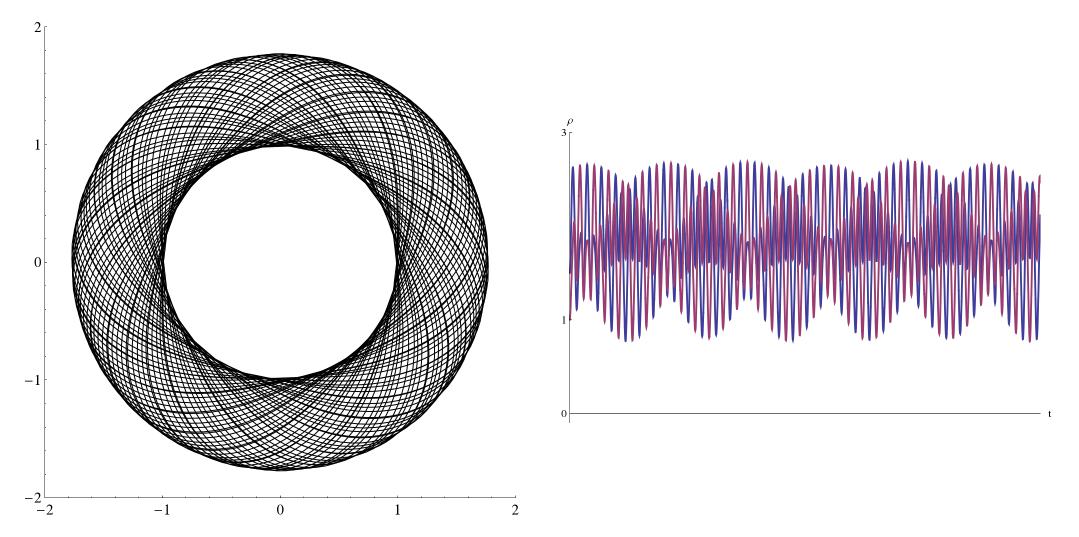}
\caption{A commutative dyonic orbit.}
\label{fig:COrb}
\end{figure}

\begin{figure}[!ht]
\cincludegraphics[width=0.9\textwidth]{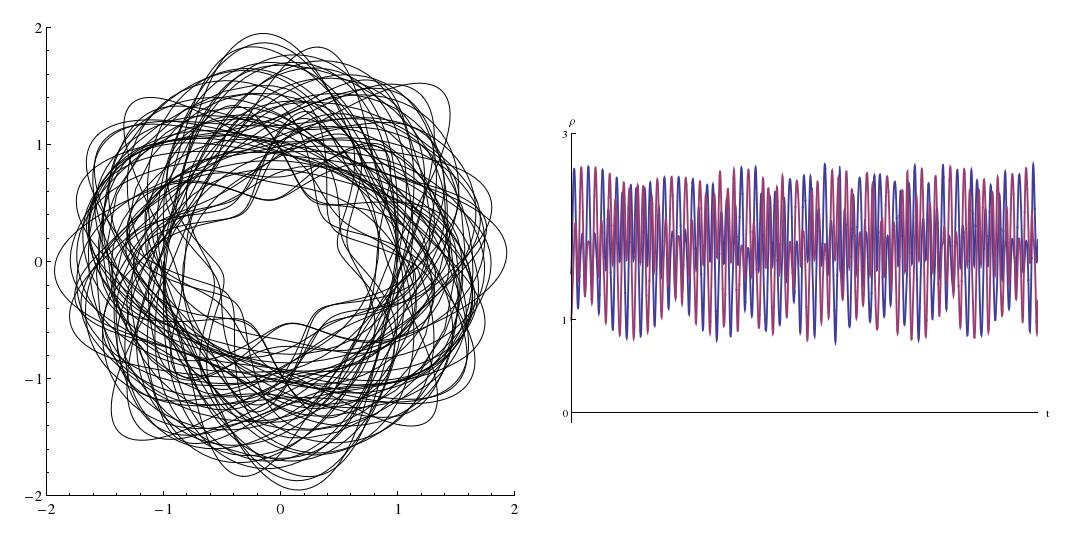}
\caption{The same initial conditions as in \figref{fig:COrb} with $\zeta=0.1$.}
\label{fig:NCOrb}
\end{figure}

The first question is whether naively adding a non-zero $\zeta$ to previously known stable orbits affects the qualitative results. We take the stable orbit previously determined and turn on some amount of noncommutativity. The differences are shown in \figref{fig:COrb} and \figref{fig:NCOrb}, where we record the evolution of the trajectories of the instantons and their sizes. In the commutative case, the instantons oscillate in a regular fashion, trading size as they sweep out a annulus in the moduli space. The maximum (minimum) combined size $\rho_{1}+\rho_{2}$ is reached on the outer (inner) edge of the annulus, as one would expect from the `free' scattering data. This orbit is stable, and exhibits no interesting features beyond those shown in \figref{fig:COrb}.

The noncommutative equivalent is less aesthetically pleasing, though it still exhibits a stable configuration. The instantons begin as in the commutative case (as can be seen most clearly in the size plots) before starting to trade sizes in an irregular fashion. This results in a more irregular orbit, but it remains stable for an indefinite period of time. The minimum distance between the two instantons is also reduced: this agrees with the results gained from the free case, where the removal of the singularity in the moduli space allows for the instantons to comfortably reside in more tightly bound configurations.

\begin{figure}[!ht]
\centering
\begin{subfigure}[t]{0.45\textwidth}
\includegraphics[width=\textwidth]{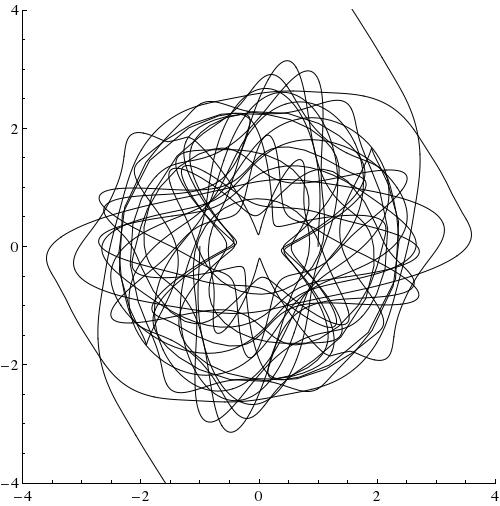}
\caption{\centering Long-lived unstable orbit for $\zeta=0.1$, $v=0.2$.}
\label{subfig:OrbEsc}
\end{subfigure}
\begin{subfigure}[t]{0.45\textwidth}
\includegraphics[width=\textwidth]{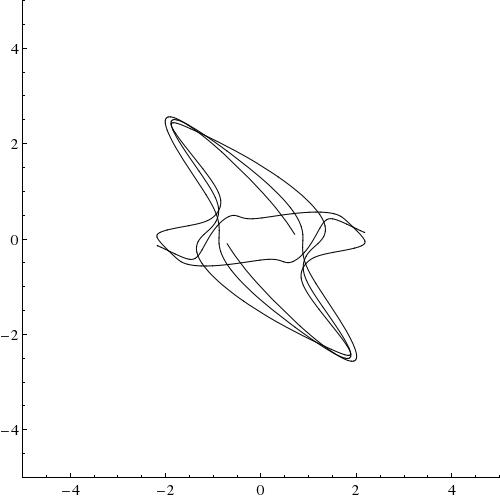}
\caption{\centering Short-lived unstable orbit for $\zeta=0.15$, $v=0.15$.}
\label{subfig:OrbAtt}
\end{subfigure}
\caption{Unstable orbit evolution.}
\label{fig:NCUnstableOrbs}
\end{figure}

Of course, this behaviour should not be assumed to be a generic feature of noncommutatively deformed orbits. As in \secref{sec:NCdyonic}, we may choose to maintain the combined effect of noncommutativity and potential, and consider the interplay between the two parameters. \figref{fig:NCUnstableOrbs} demonstrates the two configurations where, rather than remaining in a stable orbit indefinitely, the instantons attract and scatter away in finite time.

These results underline the variety of dynamical outcomes that may occur due to the presence of the additional parameter $\zeta$. It is quite probable that the set of results above is not exhaustive: it is feasible to imagine some carefully tuned system that undergoes orbit, scattering and reorbit. However, the vastness of the parameter space, coupled with the computational intensity of the numerical simulations, makes a full characterisation of the space unwieldy.

\section{Conclusion and Outlook}\label{sec:Summary}
In this work, we have calculated the metric and potential on the moduli space of two noncommutative U$(2)$ instantons using the ADHM construction. While it was not possible to find a concise form for the metric of the full, $16$-dimensional, moduli space, we were able to find a valid geodesic submanifold of the space, corresponding to non-singular fixed points of a symmetry of the metric. The key result gained is that the orbifold singularities that occur in the analogous commutative framework, due to the singular small instantons, are no longer present after the noncommutative deformation. In the large separation limit, the metric of two noncommutative instantons becomes two copies of the Eguchi-Hanson metric, in agreement with the results of \cite{Bak:2013, TongLee:2001}. Using the Manton approximation, we were able to consider scattering of slow-moving instantons, and determined that right-angled scattering is no longer a generic feature of instanton scattering, unlike in the commutative case. A variety of scattering behaviours can be demonstrated for varying noncommutative strength, including an attractive channel for pure instantons. This diversity of results extended to the dyonic picture, where previously stable orbits may have their behaviour qualitatively modified.

There are a number of open questions that remain to be addressed. While the analysis allowed for a definite parametrisation of the metric, we have only considered instanton dynamics where the the instantons have relative gauge angles in the unbroken U$(1)$ of the U$(2)$ flavour group. This gave a simple equivalence between the instanton dynamics and the vortex configurations presented in \cite{Eto:2011}, but it would be interesting (particularly in the dyonic case) to observe the effect of the full U$(2)$ on configurations. This, unfortunately, would require a full parametrisation of the ADHM data, and as such remains beyond the scope of our current work. Similarly, it would be instructive to compare the results of this moduli space approximation against the full Yang-Mills field theory. This too is beyond the reach of the computational analysis presented herein.

The aforementioned connection between instantons and vortices would also benefit from some closer consideration. It is not completely clear to what extent one can recover the non-Abelian vortex behaviour from noncommutative instantons, particularly since the identification of the instanton noncommutativity parameter $\zeta$ with the vortex U$(1)$ gauge coupling leaves us without a tuneable parameter to play the role of the vortex noncommutativity parameter. A similar connection between circle-invariant instantons and hyperbolic monopoles exists \cite{Cockburn:2014}: whether this connection can be realised in the context of noncommutative $\mathbb{R}^{4}$ remains to be seen.

It is notable that the work done in this paper is in a purely classical context. It would, therefore, be interesting to consider the corresponding quantum mechanics in a similar vein to the work of \cite{Bak:2013}. In particular, a full description of the potential term for two noncommutative U$(2)$ instantons would allow us to apply the same localisation procedure as in \cite{Stern:2000}, and gain some insight into the BPS spectra of the bound states of the quantum mechanical system. 

Finally, one could extend this work to calculate the metric of U$(N)$ noncommutative instantons for $N>2$ and $k>2$. The extension to even $N$ would perhaps be the most tractable, as the ADHM data should admit a deformed quaternion structure in the same manner as the U$(2)$ case. Such considerations would be valuable in considering the string theoretical analogue, as it would allow for the possibility of bound states that pass through D-branes and may elucidate the details of the index-counting mechanism in \cite{Stern:2000, Kim:2011} for higher gauge groups.

\acknowledgments

The authors would like to thank James Allen for his input and the use of his commutative code, and Sheng-Lan Ko for interesting discussion. A.I. would like to thank Henry Maxfield, Alex Cockburn and Paul Jennings for useful discussions, and Caroline Walters for proof-reading and feedback. A.I. is supported by an EPSRC studentship. D.J.S. is supported in part by the STFC Consolidated Grant ST/L000407/1.

\bibliographystyle{JHEP}
\bibliography{references}{}

\providecommand{\href}[2]{#2}\begingroup\raggedright\begin{thebibliography}{10}

\bibitem{Belavin:1975}
A.~A. Belavin, A.~M. Polyakov, A.~S. Schwartz, and Y.~S. Tyupkin, {\it
  Pseudoparticle solutions of the {Y}ang-{M}ills equations},  {\em Physics
  Letters B} {\bf 59} (1975), no.~1 85--87.

\bibitem{Belitsky:2000}
A.~V. Belitsky, S.~Vandoren, and P.~van Nieuwenhuizen, {\it Yang-{M}ills and
  {D}-instantons},  {\em Classical and Quantum Gravity} {\bf 17} (2000), no.~17
  3521.

\bibitem{Dorey:2002}
N.~Dorey, T.~J. Hollowood, V.~V. Khoze, and M.~P. Mattis, {\it The calculus of
  many instantons},  {\em Physics reports} {\bf 371} (2002), no.~4 231--459.

\bibitem{TongTASI:2005}
D.~Tong, {\it {TASI} lectures on solitons},  {\em arXiv preprint
  hep-th/0509216} (2005).

\bibitem{Witten:1996}
E.~Witten, {\it Small instantons in string theory},  {\em Nuclear Physics B}
  {\bf 460} (1996), no.~3 541--559.

\bibitem{Lambert:2011}
N.~Lambert, C.~Papageorgakis, and M.~Schmidt-Sommerfeld, {\it M5-branes,
  {D}4-branes and quantum 5d super-{Yang-Mills}},  {\em Journal of High Energy
  Physics} {\bf 2011} (2011), no.~1 1--17.

\bibitem{Douglas:2011}
M.~R. Douglas, {\it On $d=5$ super {Yang-Mills} theory and $(2,0)$ theory},
  {\em Journal of High Energy Physics} {\bf 2011} (2011), no.~2 1--18.

\bibitem{Stern:2000}
M.~Stern and P.~Yi, {\it Counting {Y}ang-{M}ills dyons with index theorems},
  {\em Physical Review D} {\bf 62} (2000), no.~12 125006.

\bibitem{Bak:2013}
D.~Bak and A.~Gustavsson, {\it One dyonic instanton in $5$d maximal {SYM}
  theory},  {\em Journal of High Energy Physics} {\bf 2013} (2013), no.~7
  1--52.

\bibitem{Eto:2012}
M.~Eto, T.~Fujimori, M.~Nitta, and K.~Ohashi, {\it {All Exact Solutions of
  Non-Abelian Vortices from Yang-Mills Instantons}},  {\em JHEP} {\bf 1307}
  (2013) 034, [\href{http://arxiv.org/abs/1207.5143}{{\tt arXiv:1207.5143}}].

\bibitem{Hanany:2003}
A.~Hanany and D.~Tong, {\it Vortices, instantons and branes},  {\em Journal of
  High Energy Physics} {\bf 2003} (2003), no.~07 037.

\bibitem{Manton:2014}
N.~S. Manton and P.~M. Sutcliffe, {\it Platonic hyperbolic monopoles},  {\em
  Communications in Mathematical Physics} {\bf 325} (2014), no.~3 821--845.

\bibitem{Cockburn:2014}
A.~Cockburn, {\it Symmetric hyperbolic monopoles},  {\em Journal of Physics A:
  Mathematical and Theoretical} {\bf 47} (2014), no.~39 395401.

\bibitem{Samols:1992}
T.~M. Samols, {\it Vortex scattering},  {\em Communications in Mathematical
  Physics} {\bf 145} (1992), no.~1 149--179.

\bibitem{Cho:2003}
Y.~M. Cho, H.~Khim, and N.~Yong, {\it Non-{A}belian vortices in condensed
  matter physics},  {\em arXiv preprint cond-mat/0308182} (2003).

\bibitem{Manton:1982}
N.~S. Manton, {\it A remark on the scattering of {BPS} monopoles},  {\em
  Physics Letters B} {\bf 110} (1982), no.~1 54--56.

\bibitem{Lambert:1999}
N.~D. Lambert and D.~Tong, {\it Dyonic instantons in five-dimensional gauge
  theories},  {\em Physics Letters B} {\bf 462} (1999), no.~1 89--94.

\bibitem{Peeters:2001}
K.~Peeters and M.~Zamaklar, {\it Motion on moduli spaces with potentials},
  {\em Journal of High Energy Physics} {\bf 2001} (2001), no.~12 032.

\bibitem{Nekrasov:1998}
N.~Nekrasov and A.~Schwarz, {\it Instantons on noncommutative $\mathbb{R}^{4}$,
  and $(2,0)$ superconformal six dimensional theory},  {\em Communications in
  Mathematical Physics} {\bf 198} (1998), no.~3 689--703.

\bibitem{TongLee:2001}
K.~Lee, D.~Tong, and S.~Yi, {\it Moduli space of two {U}($1$) instantons on
  noncommutative $\mathbb{R}^{4}$ and $\mathbb{R}^{3}\times \text{S}^{1}$},
  {\em Physical Review D} {\bf 63} (2001), no.~6.

\bibitem{Chu:2002}
C.~S. Chu, V.~V. Khoze, and G.~Travaglini, {\it Notes on noncommutative
  instantons},  {\em Nuclear Physics B} {\bf 621} (2002), no.~1 101--130.

\bibitem{Bruzzo:2001}
U.~Bruzzo, F.~Fucito, A.~Tanzini, and G.~Travaglini, {\it On the
  multi-instanton measure for super {Yang-Mills} theories},  {\em Nuclear
  Physics B} {\bf 611} (2001), no.~1 205--226.

\bibitem{Allen:2013}
J.~P. Allen and D.~J. Smith, {\it The low energy dynamics of charge two dyonic
  instantons},  {\em Journal of High Energy Physics} {\bf 2013} (2013), no.~2
  1--49.

\bibitem{Zwiebach:2004}
B.~Zweibach, {\em A First Course in String Theory}.
\newblock Cambridge University Press, 2004.

\bibitem{Bogomolny:1975}
E.~B. Bogomolny, {\it Stability of classical solutions},  {\em
  Sov.J.Nucl.Phys.} {\bf 24} (1976) 449.

\bibitem{ADHM:1978}
M.~F. Atiyah, N.~J. Hitchin, V.~G. Drinfeld, and Y.~I. Manin, {\it Construction
  of instantons},  {\em Physics Letters A} {\bf 65} (1978), no.~3 185--187.

\bibitem{Gopakumar:2000}
R.~Gopakumar, S.~Minwalla, and A.~Strominger, {\it Noncommutative solitons},
  {\em Journal of High Energy Physics} {\bf 2000} (2000), no.~5.

\bibitem{Hashimoto:1999}
K.~Hashimoto, H.~Hata, and S.~Moriyama, {\it Brane configuration from monopole
  solution in non-commutative super {Y}ang-{M}ills theory},  {\em Journal of
  High Energy Physics} {\bf 1999} (1999), no.~12 021.

\bibitem{MantonSutcliffe:2004}
N.~S. Manton and P.~M. Sutcliffe, {\em Topological Solitons}.
\newblock Cambridge University Press, 2004.

\bibitem{Osborn:1981}
H.~Osborn, {\it Semiclassical functional integrals for self-dual gauge fields},
   {\em Annals of Physics} {\bf 135} (1981), no.~2 373--415.

\bibitem{Kazuyuki:2001}
F.~Kazuyuki, {\it {Dp-D(p+4)} in noncommutative {Yang-Mills}},  {\em Journal of
  High Energy Physics} {\bf 2001} (2001), no.~03 033.

\bibitem{Eto:2011}
M.~Eto, T.~Fujimori, M.~Nitta, K.~Ohashi, and N.~Sakai, {\it Dynamics of
  non-{A}belian vortices},  {\em Physical Review D} {\bf 84} (2011), no.~12
  125030.

\bibitem{Eto:2006}
M.~Eto, Y.~Isozumi, M.~Nitta, K.~Ohashi, and N.~Sakai, {\it Moduli space of
  non-{A}belian vortices},  {\em Physical review letters} {\bf 96} (2006),
  no.~16 161601.

\bibitem{Manton:2010}
N.~S. Manton and N.~Sakai, {\it Maximally non-{A}belian vortices from self-dual
  {Yang-Mills} fields},  {\em Physics Letters B} {\bf 687} (2010), no.~4
  395--399.

\bibitem{Collie:2009}
B.~Collie, {\it Dyonic non-{A}belian vortices},  {\em Journal of Physics A:
  Mathematical and Theoretical} {\bf 42} (2009), no.~8 085404.

\bibitem{Kim:2011}
H.~Kim, S.~Kim, E.~Koh, K.~Lee, and S.~Lee, {\it On instantons as
  {Kaluza-Klein} modes of {M}5-branes},  {\em Journal of High Energy Physics}
  {\bf 2011} (2011), no.~12 1--52.

\end{thebibliography}\endgroup
\end{document}